\documentclass[referee]{raa}

\usepackage{graphicx,times}
\usepackage{natbib}
\usepackage{amssymb,amsmath}
\bibpunct{(}{)}{;}{a}{}{,}

\usepackage[pagebackref=true]{hyperref}

\usepackage[dvipsnames]{xcolor}
\usepackage{makecell}
\usepackage{array}

\newcommand{\msol}{\mbox{$M_\odot$}}

\newcommand{\HI}{H\,{\sc i} }
\newcommand{\hi}{\ifmmode{\rm HI}\else{H\/{\sc i}}\fi} 
\newcommand {\kms}{\ifmmode{\rm km \, s^{-1}}\else{$\rm km \, s^{-1}$}\fi} 
\newcommand{\Msun} {M_{\sun}}

\begin{document}

\title{A Catalog of Compact High-Velocity Clouds from the FAST All-Sky \HI Survey (FASHI) DR2}

\volnopage{ {\bf 20XX} Vol.\ {\bf X} No. {\bf XX}, 000--000}
\setcounter{page}{1}

\author{
Jin-Long Xu\inst{1,2}
\and Chuan-Peng Zhang\inst{1,2}
\and Xiao-Lan Liu\inst{1,2}
\and Nai-Ping Yu\inst{1,2}
\and Mei Ai\inst{1,2}
\and Ming~Zhu\inst{1,2}
}


\institute{
National Astronomical Observatories, Chinese Academy of Sciences, Beijing, 1000101, PR China
\and
Guizhou Radio Astronomical Observatory, Guizhou University, Guiyang 550000, PR China\\
}


\abstract{
High-velocity clouds (HVCs) -- especially compact HVCs (CHVCs) -- are thought to serve as gaseous tracers of dark-matter-dominated subhalos, providing a crucial empirical link between the predicted dark matter substructure and observed satellite galaxies. In this paper, we present a new catalog of 192 CHVCs identified from the FAST All-Sky \HI Survey (FASHI) Data Release 2 (DR2). Among these, 108 are ultra-compact HVCs (UCHVCs), the majority of which are new discoveries.
We find that 185 of the 192 CHVCs are spatially and kinematically concentrated around the Andromeda galaxy (M31), with a median LSR velocity of \(-296.1\ \mathrm{km\ s^{-1}}\),  suggesting physical association with M31.
Adopting an M31 distance of \(0.8\ \mathrm{Mpc}\), we derive \HI masses in the range \(1.9\times10^{5}\) to \(4.0\times10^{6}\ M_{\odot}\) and dynamical masses from \(1.0\times10^{7}\) to \(6.3\times10^{8}\ M_{\odot}\).
Approximately half of the CHVCs follow the baryonic Tully--Fisher relation.
No optical counterparts are detected in Pan-STARRS1 imaging down to a limiting magnitude of \(m_g = 22.8\).
We conclude that these CHVCs likely represent a population of gas-rich, starless minihalos and provide an excellent sample for the search for dark galaxies. 
}

\authorrunning{Xu et al.}
\titlerunning{A Catalog of Compact High-Velocity Clouds}

\maketitle

\section{Introduction}

The prevailing cold dark matter ($\Lambda$CDM) cosmological model has proven remarkably successful in reproducing the large-scale structure of the Universe. On sub-galactic scales, however, tensions persist, most famously encapsulated by the ``missing satellite problem" \citep{1999ApJ...522...82K,1999ApJ...524L..19M}. Early dark-matter-only simulations predicted that a Milky Way–sized halo should host hundreds of gravitationally bound subhalos, in stark contrast to the mere tens of luminous dwarf satellite galaxies observed at the time. Over the past two decades, this discrepancy has undergone a significant paradigm shift. High-resolution hydrodynamical simulations incorporating baryonic physics have shown that the number of luminous satellites can be largely reconciled with observations without abandoning the $\Lambda$CDM framework \citep{2022NatAs...6..897S}. In a striking reversal, deep wide-field surveys now report an unexpected overabundance of faint dwarf galaxies, potentially transforming the classical missing satellite problem into a ``too many satellites'' problem \citep{2024PASJ...76..733H,2024A&A...684L...6M}. The key insight is that inefficient star formation in low-mass halos, driven primarily by cosmic reionization and stellar feedback, renders the vast majority of subhalos optically dark or even ``born starless'' \citep{2016MNRAS.456...85S,2024ApJ...964..123J}. Nevertheless, almost all newly discovered dwarf galaxies in the Local Group are gas-poor, and gas-rich, starless low-mass halos remain conspicuously absent.

Previous models have concluded that the majority of dark matter minihalos are most likely located in the vicinity of the Milky Way, with a median distance of about 120 kpc \citep{2004ApJ...609..482K}. Within this evolving landscape, high-velocity clouds (HVCs) have emerged as a critical empirical probe. These clouds of neutral hydrogen (\HI), characterized by velocities ($V_{\rm LSR}>90$ \kms) inconsistent with a simple model of differential Galactic rotation, offer a unique window into the invisible baryon cycle and the nature of dark matter substructure \citep{1997ARA&A..35..217W}. HVCs thus represent a potential observable link between the numerous dark subhalos predicted by $\Lambda$CDM and the luminous satellite galaxies we can directly count \citep{1999ApJ...514..818B}: they can serve as gaseous tracers of otherwise invisible dark matter halos \citep{2020ApJ...900....9L}. However, the hypothesis that HVCs can fully bridge the gap between predicted subhalo populations and observed satellites is complicated by persistent observational challenges, most notably the fundamental difficulty of determining their distances and, consequently, their physical properties and gravitational binding status \citep{2012ARA&A..50..491P}.  Most strikingly, it could also be a dark-matter-dominated extragalactic system.

HVCs are often associated with more extended complexes, whereas compact high-velocity clouds (CHVCs) are small, isolated objects. Following the approach of \cite{1999ApJ...514..818B}, \cite{1999A&A...341..437B} first defined CHVCs as clouds characterised by angular sizes of less than 2$^{\circ}$, spatial isolation, and clear separation from neighbouring \HI\ emission. \cite{2002ApJS..143..419S} subsequently modelled the physical state of \HI\ gas confined within dark matter minihalos and concluded that the observed CHVC parameters are consistent with a circumgalactic population located at typical distances of 150 kpc. CHVCs are therefore regarded as ideal candidates for identifying gas-rich, low-mass dark halos. Over the years, several surveys with different instruments have been employed to search for CHVCs. \cite{1999A&A...341..437B} used the \HI\ Leiden/Dwingeloo Survey (LDS) to compile a catalogue of 66 CHVCs. An improved catalogue containing 67 objects was subsequently published by \cite{2002A&A...391..159D} based on the same survey, using an automated search routine. In addition, \cite{2002AJ....123..873P} extracted 179 CHVCs from the \HI\ Parkes All-Sky Survey (HIPASS). More recently, \cite{2012ApJ...758...44S} presented a catalogue of 1964 CHVCs identified in the Galactic Arecibo L-Band Feed Array (ALFALFA) survey. In addition, ultra-compact high-velocity clouds (UCHVCs) were defined by \cite{2013ApJ...768...77A} as HVCs with a local standard of rest velocity $\mid V_{\rm LSR} \mid$  $>$ 120 \kms, an \HI\ major axis size $<30^{\prime}$, and a signal-to-noise ratio (SNR) $>$ 8. Using the 40\% complete ALFALFA \HI-line survey, they compiled a catalogue of 59 UCHVCs and inferred that these objects are consistent with being very low-mass galaxies within the Local Volume.

Several CHVCs have been studied in great detail \citep{2004A&A...426L...9B,2005A&A...432..937W,2014A&A...563A..99F}. A particularly striking piece of evidence for this formation channel was recently reported by \cite{2025SciA...11S4057L}, who identified a CHVC, AC G185.0-11.5, located within the HVC complex AC-I. Using the baryonic Tully-Fisher relation, they estimated its distance to be approximately 277.7 kpc, placing it firmly in the Local Group and well beyond the disk of the Milky Way. Critically, this object exhibits kinematic signatures consistent with a rotating disk galaxy, yet it lacks any detectable stellar counterpart. This suggests that it is a rare dark galaxy — a gas-rich dark matter halo that has failed to form stars. The subsequent discovery of Leo P from ALFALFA survey data demonstrated that galaxies similar to Leo T indeed exist in the Local Volume \citep{2013AJ....146...15G,2013AJ....145..149R,2013AJ....146....3S}. Both Leo P and Leo T are considered gas-rich ultra-faint dwarf galaxies. Notably, Leo P was discovered during the routine identification of \HI\ detections within the ALFALFA survey, when it was noticed that one UCHVC could be associated with an irregular, lumpy light distribution in SDSS images \citep{2013AJ....146...15G}. More recently, \cite{2025ApJ...982L..36X} identified a UCHVC with a systemic velocity of 127.5 \kms, which coincides in position with KK 153 based on optical observations. KK 153 is forming stars in a cold-gas-dominated disk and is likely a gas-rich ultra-faint dwarf galaxy.

In this paper, we present a new catalog of CHVCs and UCHVCs detected by the FAST All-Sky \HI Survey (FASHI).
We describe the selection method, analyze the spatial and kinematic distributions, and investigate the physical properties of these clouds, including their \HI masses, dynamical masses, and optical counterparts.

\section{Observations}
\label{sec:obs}
\subsection{The FAST All Sky \HI Survey (FASHI)}
The FAST All Sky \HI Survey (FASHI) conducted with the Five-hundred-meter Aperture Spherical radio Telescope (FAST).
FASHI survey is designed to map the entire sky accessible to FAST, covering declinations from -14$^{\circ}$ to +66$^{\circ}$ in the frequency range 1.0–1.5 GHz. Between August 2020 and June 2023, the FASHI DR1 spans more than 7600 square degrees and detected  a total of 41741 extragalaxctic \HI sources \citep{2024SCPMA..6719511Z}. Between August 2020 and July 2025, FASHI observed approximately 19500 square degrees, corresponding to 47.3\% of the full sky. Within this area, the survey has identified about 156 000 extragalactic \HI sources at redshifts z$<$0.09 \cite[DR2;][]{2026arXiv260631539Z}.

FAST is located in Guizhou, China, and is currently the most sensitive single-dish radio telescope in the world \citep{2019SCPMA..6259502J,2020RAA....20...64J}. FAST observations were performed using drift-scan and multi-beam on-the-fly (OTF) modes, with 19 beams simultaneously.
The scan velocity was set to \(15''\ \mathrm{s^{-1}}\) with an integration time of 1 second per spectrum.
The digital backend provides a bandwidth of 500 MHz with 64k channels, yielding a frequency resolution of \(7.629\ \mathrm{kHz}\) (velocity resolution of \(1.6\ \mathrm{km\ s^{-1}}\)).
The half-power beam width (HPBW) is \(\sim2.9'\) at 1.4 GHz, and the pointing accuracy is better than \(10''\).
The system temperature during observations was \(16\)--\(22\ \mathrm{K}\).
Data reduction was performed using the Python-based pipeline HIFAST \citep{2024SCPMA..6759514J}.
The final data cubes have a pixel size of \(1.0' \times 1.0'\), and the velocity resolution was smoothed to \(6.4\ \mathrm{km\ s^{-1}}\).
The median sensitivity is \(0.57\ \mathrm{mJy\ beam^{-1}}\).

\subsection{The Pan-STARRS1 Survey}
Optical imaging data are taken from the Pan-STARRS1 (PS1) survey \citep{Chambers+2016}.
PS1 covers three-quarters of the northern sky (declinations \(>-30^\circ\)) with a 1.8-meter telescope.
For the stacked \(3\pi\) Steradian Survey, the \(5\sigma\) point-source limiting magnitudes are 23.3 (\(\rm g\)), 23.2 (\(\rm r\)), 23.1 (\(\rm i\)), 22.3 (\(\rm z\)), and 21.4 (\(\rm y\)).
In this study, we mainly use \(\rm g\)-band and \(\rm r\)-band images.

For images where no optical counterpart is detected, we generally need to provide the detection limit magnitude of the image. However, typical optical surveys only provide the limiting sensitivity of point sources. Based on the limiting sensitivity of the point source, we can estimate the limiting sensitivity of the image. The apparent magnitude for the g-band image can be determined by 
\begin{equation}
      m_{\rm g}=m_{\rm zp}-2.5 \times\rm log(\it F)
\end{equation}
where $m_{\rm zp}$ is zero magnitude, and $F$ is total image flux. Hence, the limiting apparent magnitude ($m_{\rm g}^{i}$) of the image can be obtained by
\begin{equation}
      m_{\rm g}^{i}= m_{\rm g}^{p}+2.5\times \rm log(\it F_{p}/F_{i})
\end{equation}
where $m_{\rm g}^{p}$ is the g-band point-source limiting apparent magnitude, and the total flow ratio of point source and image can be calculated by
\begin{equation}
      \dfrac{F_{p}}{F_{i}}=\dfrac{M\times \rm rms\times\sqrt{n_{p}}}{N\times \rm rms\times\sqrt{n_{i}}}=\dfrac{M\times\sqrt{n_{p}}}{N\times \sqrt{n_{i}}}
\end{equation}
where $M$ is the signal-to-noise ratio taken from the point source, while $N$ is the signal-to-noise ratio taken from the image. $n_{i}$ is the pixels occupied by the photometry area of the image, which can be determined by $n_{i}=S_{\rm size}/(\it P^{2})$, where $S_{\rm size}$ is photometry area and $P$ is pixel size. For the point source,  $n_{p}=\pi\times(\rm FWHM/2)^{2}/(\it P^{2})$, where FWHM is the Full Width at Half Maximum of the point spread function (PSF). Hence, we can obtain
\begin{equation}
     \dfrac{n_{p}}{n_{i}}=\dfrac{\pi\times(\rm FWHM/2)^{2}}{S_{\rm size}}
\end{equation}
For the Pan-STARRS1 (PS1) survey, FWHM is 1.47$^{\prime\prime}$ in g-band. The 5$\sigma$ point-source limiting apparent magnitude is 23.3, so the $M$ can be adopted as 5. For images without detected optical sources, the sky background signal dominates the flux measurement \citep{2021AJ....162..274L}. We therefore adopt the $1\sigma$ flux distribution within $10^{\prime\prime} \times 10^{\prime\prime}$ regions to determine the upper limits on the absolute magnitude for image. Here we can take $N$ as 1. Therefore, we derived that $m_{\rm g}^{i}$ is 22.8 magnitude in the g-band image. Similarly, we also obtained that $m_{\rm r}^{i}$ is 22.6 magnitude in the r-band image.

\begin{figure*}
\centering
\includegraphics[width=0.95\textwidth]{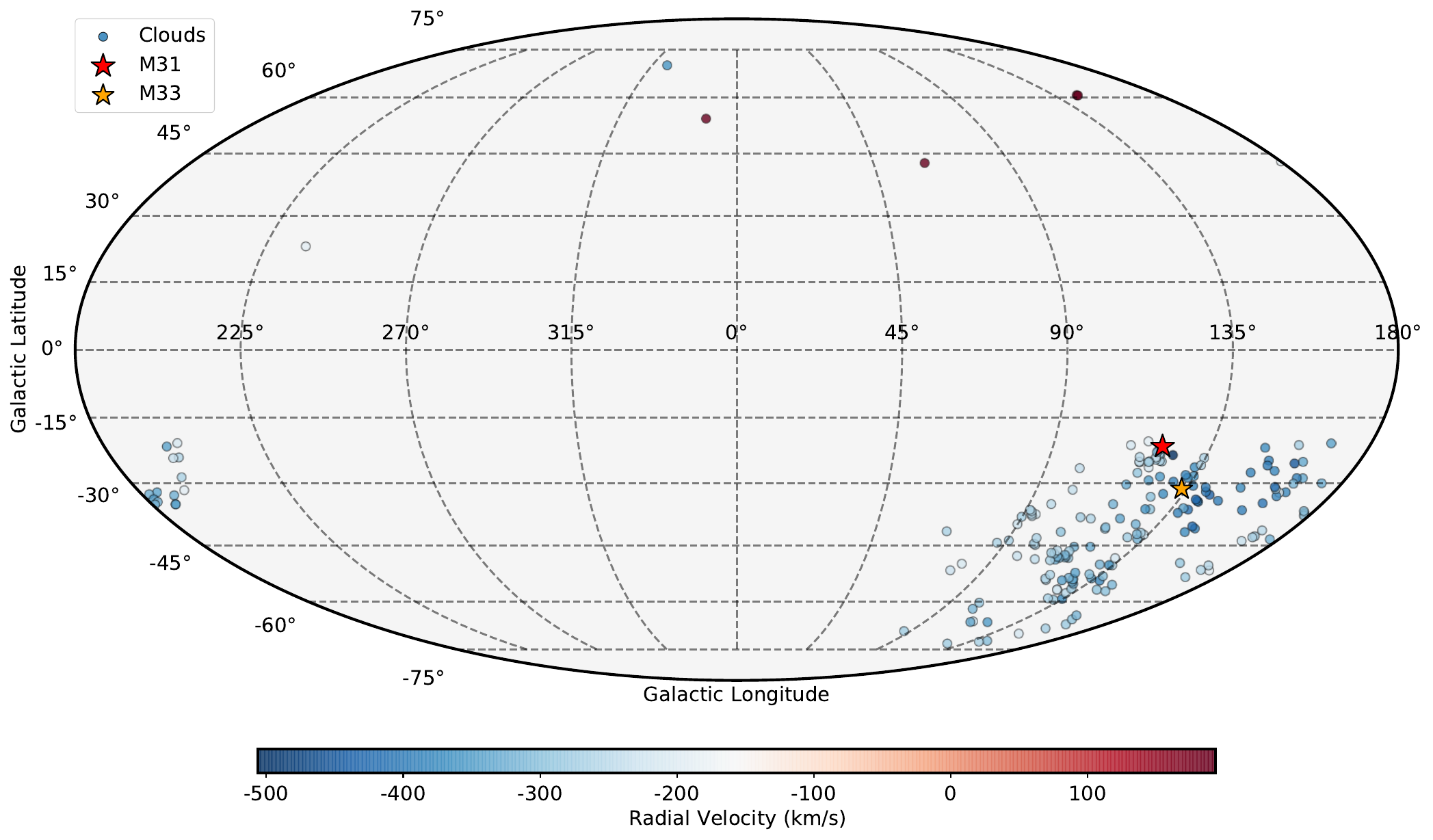}
\vspace{-3mm}
\caption{The spatial distribution of CHVCs selected from the FASHI DR2 source catalog. As a comparison, we also drew the positions of M31 and M33.}
\label{fig:CHVCs_all}
\end{figure*}

\section{Results}
\label{sect:results}
\subsection{Selection of the CHVC Sample}
The FASHI DR2 survey has identified approximately 156,000 extragalactic \HI\ sources at redshifts z$<$0.09 \citep{2026arXiv260631539Z}. The DR2 source catalogue not only fully encompasses the DR1 coverage but also includes the ALFALFA survey region. While DR1 was limited to sources with velocities greater than 200 \kms, DR2 extends the catalogue to include lower-velocity sources. To identify these \HI\ sources, the FASHI team employed the Source Finding Application (SoFiA) on the \HI\ data cubes \citep{2021MNRAS.506.3962W}. Because all sources were extracted using a unified, standardised procedure for galaxy searching, the automated source finder ensures that each detected source is locally isolated and independent of extended \HI\ emission structures. These identified \HI\ sources are likely to be associated with galaxies, starless cold gas clouds, and HVCs. Using the DR1 source catalogue, \cite{2026MNRAS.548ag732M} identified 70 dark galaxy candidates within 50 Mpc, excluding the Local Volume (within 11 Mpc). CHVCs and dark galaxies are very similar, as both are starless and gas-rich. In this work, we used the FASHI DR2 source catalogue to select our CHVC sample.

To select the CHVC sample in the Local Group, we restricted the source velocities to less than 200 \kms. This also ensures that the selected CHVCs do not overlap with the dark galaxy candidates studied by \cite{2026MNRAS.548ag732M}. We identified a total of 347 \HI\ sources, including some optically bright galaxies. To exclude these galaxies, we cross-matched our sample with optical data from the Pan-STARRS1 survey and the NASA/IPAC Extragalactic Database (NED). For each source, we searched for optical counterparts within a circular aperture of 3 arcmin in diameter, based on its coordinate position and radial velocity, and checked for corresponding names in the databases. In the region covered by the Dark Energy Spectroscopic Instrument (DESI), some sources lacked optical counterparts in the available surveys, so we inspected the DESI images directly.  In addition, to minimise the impact of interstellar extinction on our classification, we restricted the sample to sources at Galactic latitudes outside the range -20$^{\circ}$ to 20$^{\circ}$. Finally, by adopting the criteria that a CHVC should have an angular size of less than 2$^{\circ}$ and a local standard of rest velocity $| V_{\rm LSR}|>$90 \kms, we selected a total of 192 CHVCs. Throughout this work, we adopt the major axis diameter ($D_{\rm maj}$)  of the best-fitting ellipse to each cloud as its size from the FASHI DR2 catalog.

Previous studies have compiled CHVC catalogues from various \HI\ surveys. \cite{1999A&A...341..437B} used the \HI\ LDS to compile a catalogue of 66 CHVCs, while \cite{2002AJ....123..873P} extracted 179 CHVCs from the HIPASS. More recently, \cite{2012ApJ...758...44S} presented a catalogue of 1964 CHVCs from the ALFALFA survey. To cross-match our sample with these existing catalogues, we adopted a distance threshold equal to the FASHI beam size and a velocity threshold equal to twice the FASHI velocity resolution. Under these criteria, we found that only 8 of our CHVCs had been previously detected. Figure \ref{fig:CHVCs_all} shows the spatial distribution of our CHVC sample. We find that the majority of the CHVCs are located in the vicinity of M31 and M33, with only 7 situated in the northern sky region. In addition, \cite{2005A&A...436..101W} identified 17 HVCs around M31 and near M33. Among these, only one of our CHVCs - CHVC120.74-23.57-506s - coincides with their sample.

\begin{table*} 
     \scriptsize{
     \centering
      \caption{Measured and derived properties of CHVCs.}
      \vspace{-8pt}
      \setlength{\tabcolsep}{1.5pt}
      \label{tab:Obs}
      \begin{tabular}{lcccccccccccc}
\noalign{\vspace{0pt}}\hline\hline\noalign{\vspace{0pt}}
Name & FASHI-id &RA+Dec. & $V_{\rm LSR}$ & $D_{\rm maj}*\it D_{\rm min}$ & $W_{50}$  & $S_{\rm sum}$& SNR  & log($N_{\rm HI}$)&  $R_{\rm eff}$ &log$M_{\rm HI}$ & log$M_{\rm dyn}$\\
 CHVC+l+b+Vsys&  & [J2000]  & [\kms] &[$\prime\times\prime$]  &  [\kms] &  [mJy$\cdot$ km s$^{-1}$] & & [cm$^{-2}]$ &[kpc] & [$\msol$]&[$\msol$]\\
\noalign{\vspace{0pt}}\hline\noalign{\vspace{0pt}}
CHVC99.20-57.03-299 & 20260000068 & 000044+0329 & -298.8 & 45.0$\times$36.2 & 75.7 & 6.7 & 20.9 & 19.2 & 1.9 & 6.0 & 8.8\\
CHVC100.79-55.27-318 & 20260000175 & 000136+0526 & -317.6 & 50.8$\times$28.8 & 31.4 & 5.3 & 30.5 & 19.4 & 1.8 & 5.9 & 8.0\\
CHVC111.35-28.42-360 & 20260000272 & 000238+3321 & -359.8 & 89.0$\times$28.2 & 27.7 & 4.2 & 18.7 & 19.5 & 2.3 & 5.8 & 8.0\\
CHVC89.92-67.99-274 & 20260000423 & 000348-0813 & -274.3 & 23.4$\times$13.4 & 19.0 & 1.3 & 15.0 & 19.2 & 0.8 & 5.3 & 7.2\\
CHVC112.65-26.36-368 & 20260000673 & 000603+3535 & -367.5 & 103.2$\times$49.8 & 25.5 & 13.2 & 19.6 & 20.0 & 3.3 & 6.3 & 8.1\\
CHVC112.88-28.83-368 & 20260001005 & 000918+3313 & -368.5 & 17.6$\times$12.0 & 25.3 & 0.5 & 10.4 & 18.7 & 0.7 & 4.9 & 7.4\\
CHVC103.60-56.62-296 & 20260001006 & 000919+0439 & -295.6 & 29.2$\times$16.6 & 29.2 & 1.7 & 20.6 & 19.1 & 1.0 & 5.4 & 7.7\\
CHVC113.98-28.14-384 & 20260001447 & 001317+3403 & -383.7 & 27.2$\times$25.2 & 27.3 & 1.7 & 17.2 & 19.1 & 1.2 & 5.4 & 7.7\\
CHVC96.43-69.64-217 & 20260001660 & 001519-0839 & -216.6 & 40.0$\times$30.2 & 24.9 & 6.7 & 29.6 & 19.6 & 1.6 & 6.0 & 7.8\\
CHVC108.05-54.14-385 & 20260001798 & 001630+0742 & -385.3 & 44.4$\times$30.0 & 26.7 & 2.7 & 22.4 & 19.2 & 1.7 & 5.6 & 7.8\\
\noalign{\vspace{0pt}}\hline\hline\noalign{\vspace{6pt}}
\end{tabular}
\\Note: Please refer to the appendix for the remaining tables. ``S" represent that this CHVC is associated with \cite{2012ApJ...758...44S}; ``W" for \cite{2005A&A...436..101W}; ``A" for \cite{2013ApJ...768...77A}}
\end{table*}

\subsection{The CHVC Catalog}
Our catalog of 192 CHVCs is shown in Table 1 and Appendix A. Some parameters are directly taken from the FASHI DR2 catalog, while others are obtained through calculation. The columns of the table are defined in the following way.

Column 1: The CHVC's name, which is defined as the integer rounded Galactic longitude and latitude together with the  respect to the Local Standard of Rest  (LSR) velocity.

Column 2:  The identification (ID) number of each CHVC in the FASHI catalog. 

Columns 3 and 4: J2000 right ascension and declination coordinates of the CHVC centroid.

Column 5:  The radial velocity  ($V_{\rm LSR}$). In the FASHI catalog, the radial velocity is displayed as Heli-centric velocity ($V_{\rm Hel}$). To align with the definition of HVCs, we have transformed it into $V_{\rm LSR}$.

Columns 6 and 7:  The major and minor axes for an ellipse, which represents the measurement aperture of the intergrated source for estimating the source flux. 

Column 8:  $W_{50}$ is  the velocity width of the \HI line profile of each CHVC, which is measured at 50\% level of every peak by busy-function fitting.

Column 9:  Integrated \HI line flux density ($S_{\rm sum}$), which sums all the velocity channels containing sigal for each integrated spectrum.

Column 10:  Signal-to-noise (SNR) of the detection for each CHVC.

Column 11:  Peak column density, estimated as $N_{\rm HI} = (1.1\times10^{24}/B_{\rm s}^{2})\times F_{\rm peak}\times dv$,  where $dv$ is the velocity width in \kms (6.4 \kms), $B_{\rm s}$ is the width of beam in arcsec (174$^{\prime}$), and $F_{\rm peak}$ is  peak flux density in Jy/beam of each integrated spectrum for each CHVC.

Column 12:  Effective radius, which can be determined by  $R_{\rm eff} =D_{\rm dis} \sqrt{(D_{\rm maj}/2.5)\times (D_{\rm min}/2.5)}/2$, where $D_{\rm dis}$ is the distance to each CHVC, and $D_{\rm maj}$ and $D_{\rm min}$ are the major and minor axes for an fitting ellipse  of each CHVC. For the FASHI catalog, the fitting ellipse is set to be approximately 2.0-2.5 times larger than the area covered by the 4.5$\sigma$ region. We will divide the the major and minor axes by 2.5. Hence, we have obtained a minimum effective radius here.

Columns 13 and 14:  \HI mass and dynamic mass. The \HI gas mass of each CHVC can be calculated as $M_{\rm HI} =  2.36\times10^{5}D^{2}_{\rm dis}S_{\rm sum}$. Moreover, the dynamical mass can then be computed as $M_{\rm dyn} = V_{\rm rot}^{2} R_{\rm HI} / G$, $R_{\rm eff}$ is the effective radius from Column 12, $G$ is the gravitational constant, and $V_{\rm rot}$ is rotational velocity.

\begin{figure*}
\centering
\includegraphics[width=0.95\textwidth]{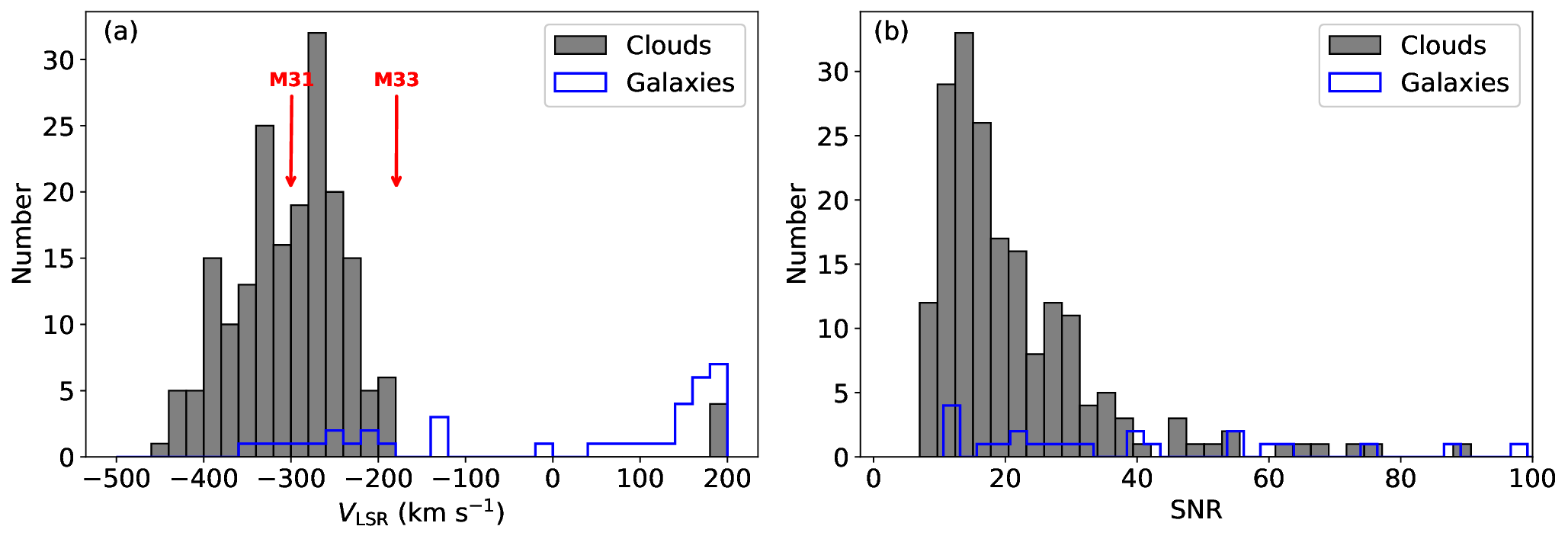}
\includegraphics[width=0.95\textwidth]{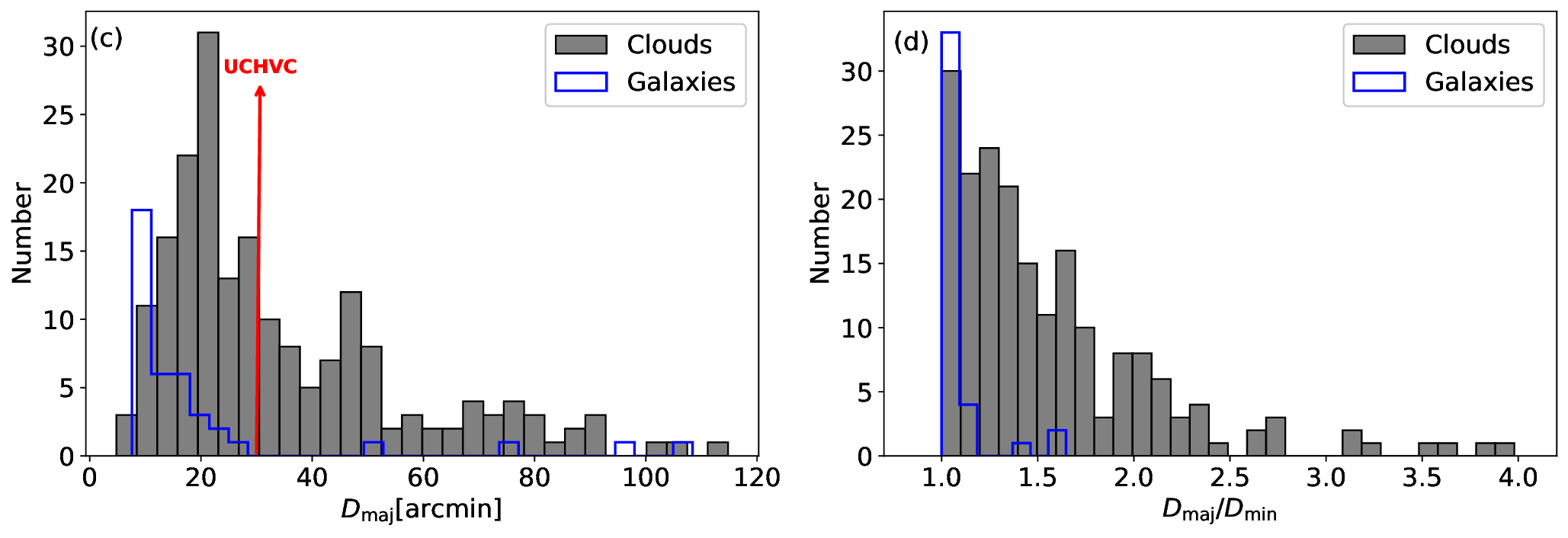}
\includegraphics[width=0.95\textwidth]{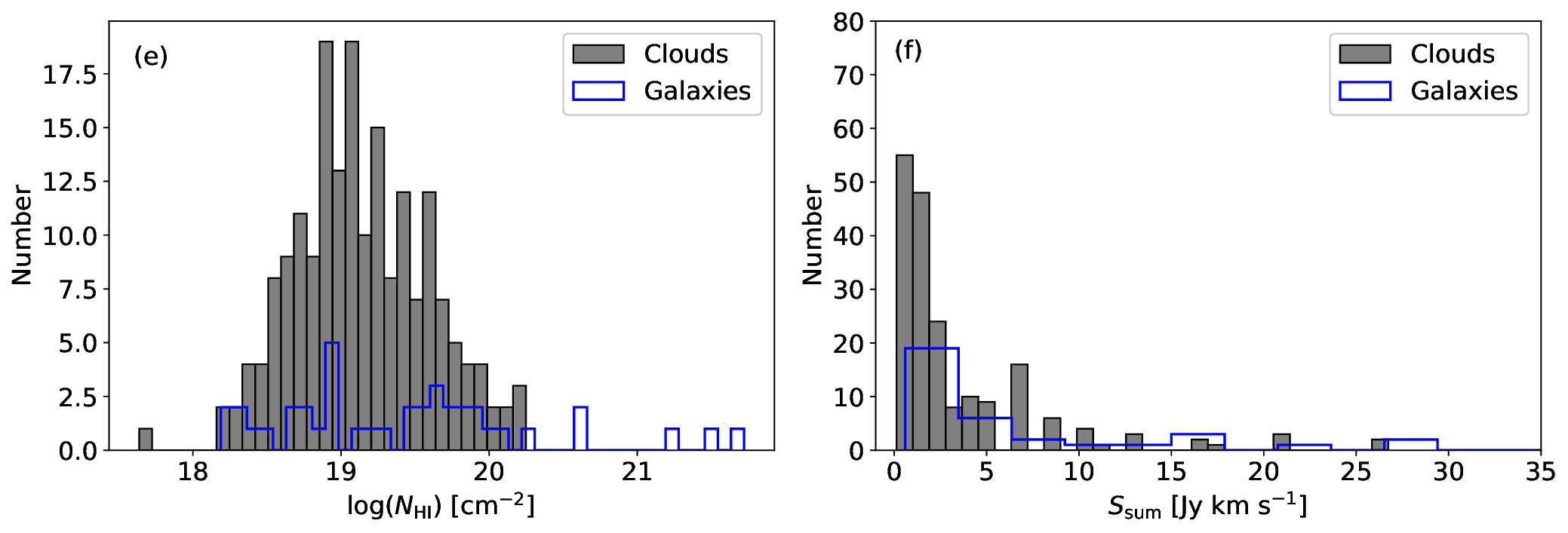}
\includegraphics[width=0.95\textwidth]{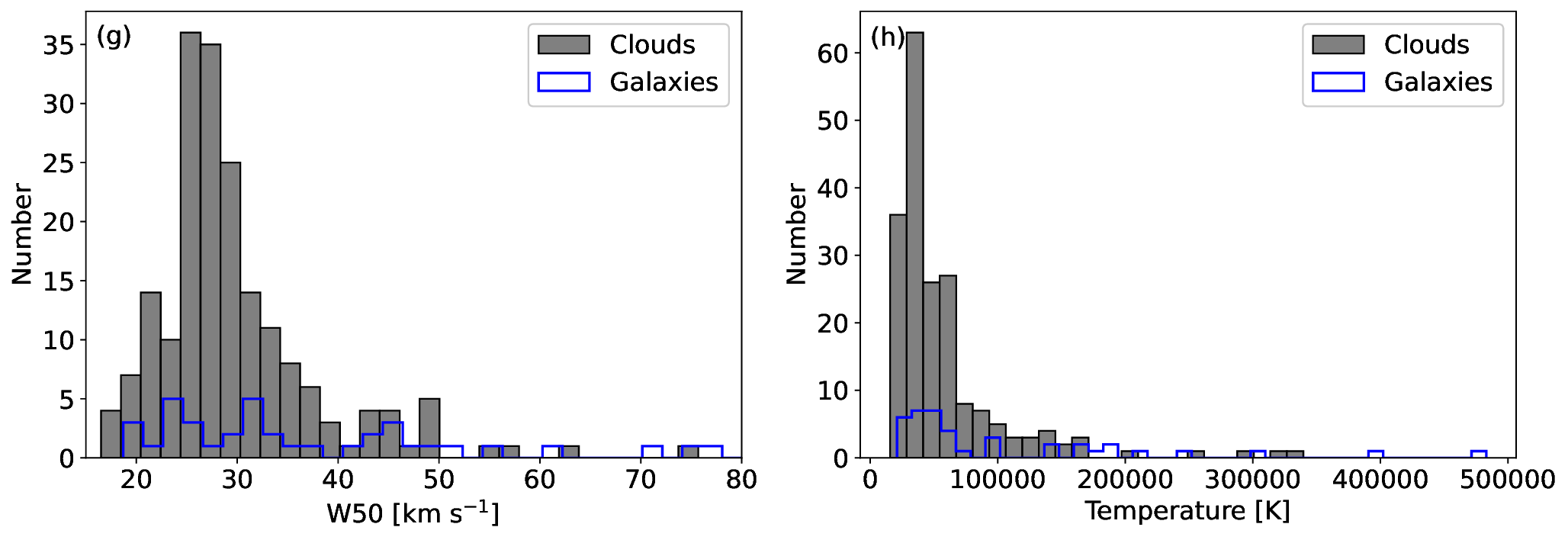}
\vspace{-3mm}
\caption{Histograms of measured properties for the CHVCs. (a) system velocity. (b) signal-to-noise ratio. (c) major axis. (d) ratio of major axis to minor axis. (e) column density. }
\label{fig:Histograms}
\end{figure*}

\subsection{Basic Properties of the CHVC Sample}
In addition to the 192 CHVCs, our selection also yielded a number of known galaxies, including some intermediate-mass and high-mass systems. To facilitate comparison with the CHVC sample, we removed galaxies with masses greater than 10$^{8}$ $\Msun$, retaining 40 low-mass dwarf galaxies.  Figure \ref{fig:Histograms} presents histograms of the measured properties for both the selected CHVCs and the dwarf galaxy sample. We find that 188 of the 192 CHVCs have $V_{\rm LSR}$ values between -506.2 \kms and -191.5 \kms, while only 4 CHVCs exhibit larger than 120 \kms.  As shown in the histogram in Figure \ref{fig:Histograms}a, the $V_{\rm LSR}$ velocities of these 188 CHVCs are not randomly distributed, in contrast to the 40 dwarf galaxies, but are instead strongly concentrated.
The radial velocities of M31 and M33 are -300.0 \kms and -179.2 \kms \citep{2012AJ....144....4M}, respectively. We obtained that the median value of $V_{\rm LSR}$ for these 188 CHVC is -296.1 \kms, which is just associated with radial velocity of M31, as also shown in Figue \ref{fig:Histograms}a. This strong kinematic coherence confirms that most CHVCs are physically associated with M31, not the Milky Way.

The signal-to-noise ratios (SNR) of these 192 CHVCs range from 7.0 to 195.9, with a median value of 17.5. As shown in Figure \ref{fig:Histograms}b, the SNR values of most CHVCs are concentrated between 7 and 40. The major axes ($D_{\rm maj}$) of these CHVCs range from 4.8 to 114.8, with a median value of 27.1, whereas those of the selected dwarf galaxies range from 7.6 to 108.4, with a median value of 11.7. \cite{2013ApJ...768...77A} defined ultra-compact high-velocity clouds (UCHVCs) as HVCs with $\mid V_{\rm LSR} \mid$  $>$ 120 \kms, \HI major axis size $<30^{\prime}$, and SNR $>$ 8. Using the 40\% complete ALFALFA \HI-line survey, they compiled a catalog of 59 UCHVCs. Based on this definition, 108 of our 192 CHVCs qualify as UCHVCs. Only 2 of these CHVCs are included in the UCHVC catalog of \cite{2013ApJ...768...77A}; the remaining 106 are newly discovered. Interestingly, 90\% of the low-mass dwarf galaxies also satisfy the UCHVC definition, suggesting that UCHVCs are excellent dwarf galaxy candidates, as also shown in Figure \ref{fig:Histograms}c. This suggests that UCHVCs can be considered as dwarf galaxy candidates. From Figure \ref{fig:Histograms}d, we see that the $D_{\rm maj}$/$D_{\rm min}$  values of  all dwarf galaxies are smaller than 2, and mainly concentrated between 1 and 1.2.
For the CHVCs, the $D_{\rm maj}$/$D_{\rm min}$ values range from 1.0 and  4.0, with a median value of 1.4. If a ratio greater than 3 is defined as indicative of a filamentary structure, then most of our CHVCs, similar to the dwarf galaxies, exhibit a compact morphology.

Figues \ref{fig:Histograms}e and \ref{fig:Histograms}f show the histograms of peak \HI column density ($N_{\rm HI}$) and total flow density ($S_{\rm sum}$), respectively. The distribution of $N_{\rm HI}$ and $S_{\rm sum}$ are similar for the CHVCs and the dwarf galaxies.  Their log($N_{\rm HI}$) values are mainly concentrated between 18.2 and 20.3, and their $S_{\rm sum}$ values are smaller than 11 Jy \kms. For the CHVCs, the median $N_{\rm HI}$ is 1.2$\times$10$^{19}$ cm$^{-2}$, which is consistent with the values found for CHVCs from LDS and HIPASS \citep{2002A&A...391..159D,2002AJ....123..873P}. The median $S_{\rm sum}$ is 1.7 Jy \kms, close to that (1.3 Jy \kms) of the gas-rich ultra-faint dwarf galaxy Leo P \citep{2013AJ....146...15G,2013AJ....145..149R,2013AJ....146....3S}. Figue \ref{fig:Histograms}g shows the histograms of $W_{50}$ and temperature. The $W_{50}$ of these CHVCs range from 16.5 \kms and  75.7 \kms,  with a median value of 27.7 \kms. Such a median value of $W_{50}$ is  associated with that of CHVCs from LDS (25 \kms),  and is slightly lower than that from HIPASS (35 \kms). The velocity dispersion $\sigma = W_{50}/\sqrt{8\rm\ln(2)}$. Based on the velocity dispersion, we derive the kinetic temperature of these CHVCs using $ T = m_{\rm H}\sigma^{2}/k$, where $m_{\rm H}$ is hydrogen atomic mass and $k$ is boltzmann constant. Figure \ref{fig:Histograms}h shows the histogram of temperature. In the \HI gas, the temperature of a cool neutral medium (CNM) component is generally $\leq$1000 K, while that of a warm neutral medium (WNM) component is $\geq$5000 K \citep{1996ApJ...462..203Y,2018A&A...612A..26A}. We find that the temperatures of all the CHVCs exceed 5000 K, indicating that all the CHVCs are warm. 

\begin{figure*}
\centering
\includegraphics[width=0.9\textwidth]{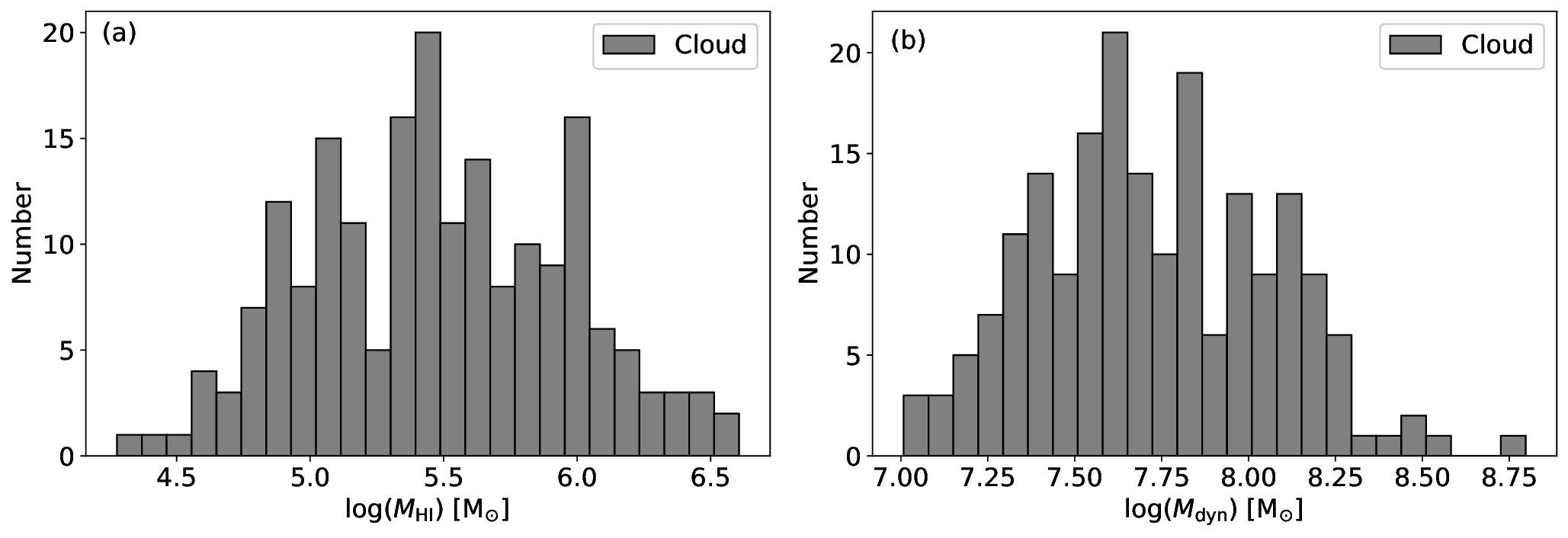}
\vspace{-3mm}
\caption{Histograms of obtained masses for the CHVCs. (a) \HI mass. (b) dynamic mass.}
\label{fig:mass}
\end{figure*}

\section{DISCUSSION}
From the FASHI DR2 catalog, we selected 192 CHVCs. Among them, 185 objects located in the southern sky are spatially projected near M31 and M33. Based on statistical and theoretical arguments from O VI (O$^{5+}$) absorbers detected in active galactic nuclei, CHVCs have been proposed to reside at large characteristic distances ($>$100 kpc) within the Local Group (LG) \citep{2003Natur.421..719N}. As shown in Figure \ref{fig:CHVCs_all}, all 185 spatially clustered CHVCs exhibit negative radial velocities. The $V_{\rm LSR}$ histogram further reveals that these negative-velocity CHVCs are kinematically associated with M31. Their positional and velocity distributions suggest that they form a clustered population around M31 and likely share its distance of 0.8 Mpc \citep{2012AJ....144....4M}. Only seven CHVCs are located in the northern sky region and lack distance constraints; we verified that excluding them does not alter the statistical distributions in Figure \ref{fig:Histograms}. We therefore focus the following discussion on the properties of these 185 CHVCs.

\begin{figure}
\centering
\includegraphics[width=0.47\textwidth]{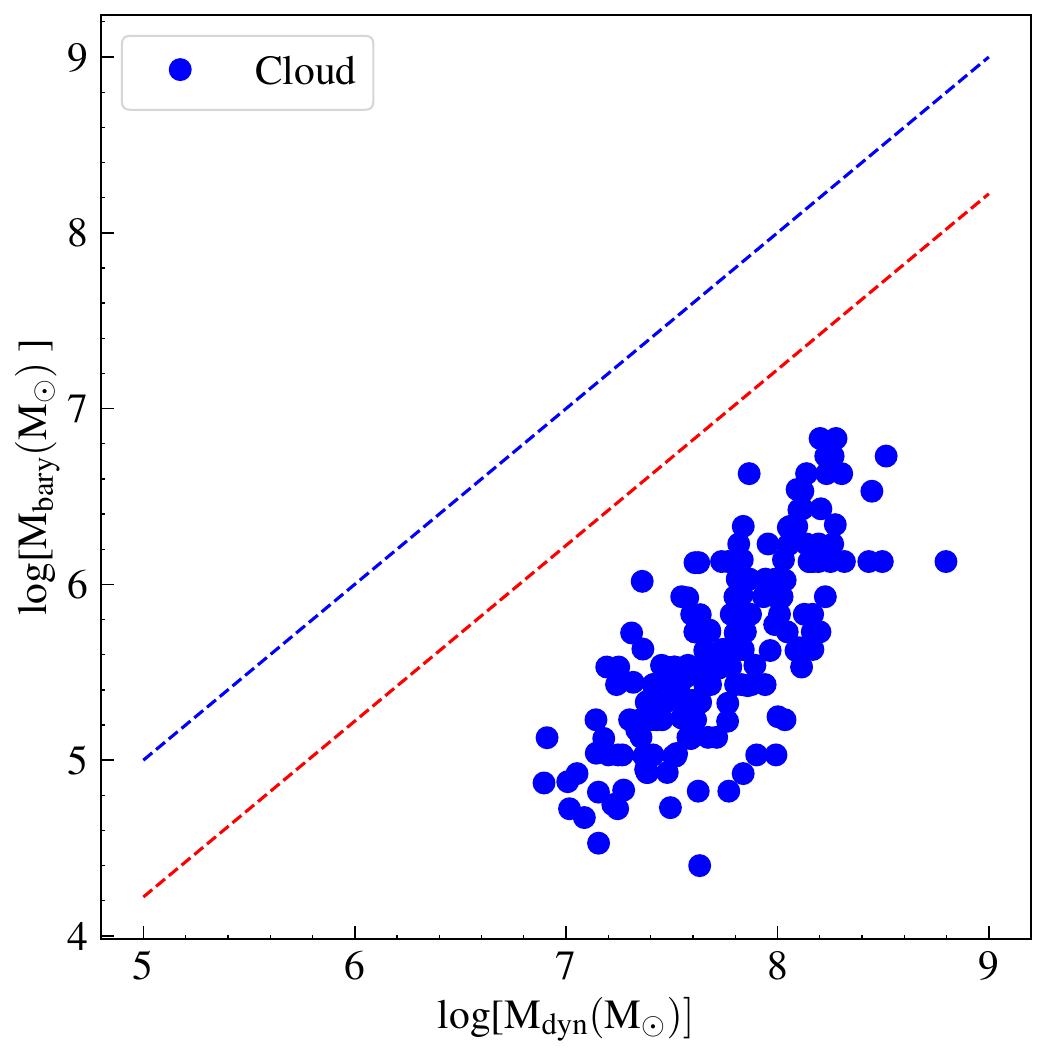}
\vspace{-3mm}
\caption{The relationship between dynamic mass ($M_{\rm dyn}$) and baryon mass ($M_{\rm bary}$). The  blue dotted line indicates where $M_{\rm dyn}$ equals $M_{\rm bary}$, while the red dotted line marks $M_{\rm dyn}$=6$M_{\rm bary}$.}
\label{fig:dyn_DV}
\end{figure}

\begin{figure}
\centering
\includegraphics[width=0.47\textwidth]{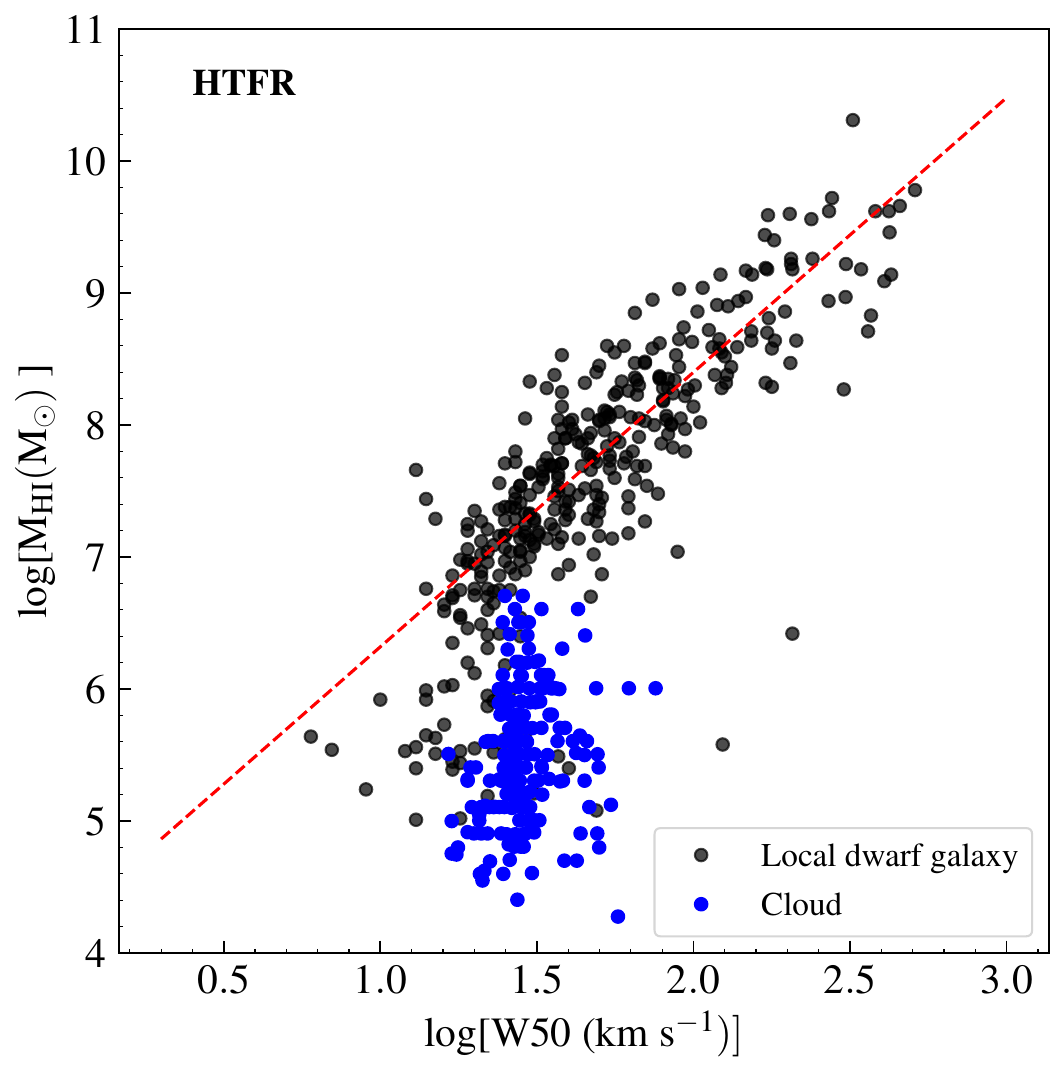}
\includegraphics[width=0.47\textwidth]{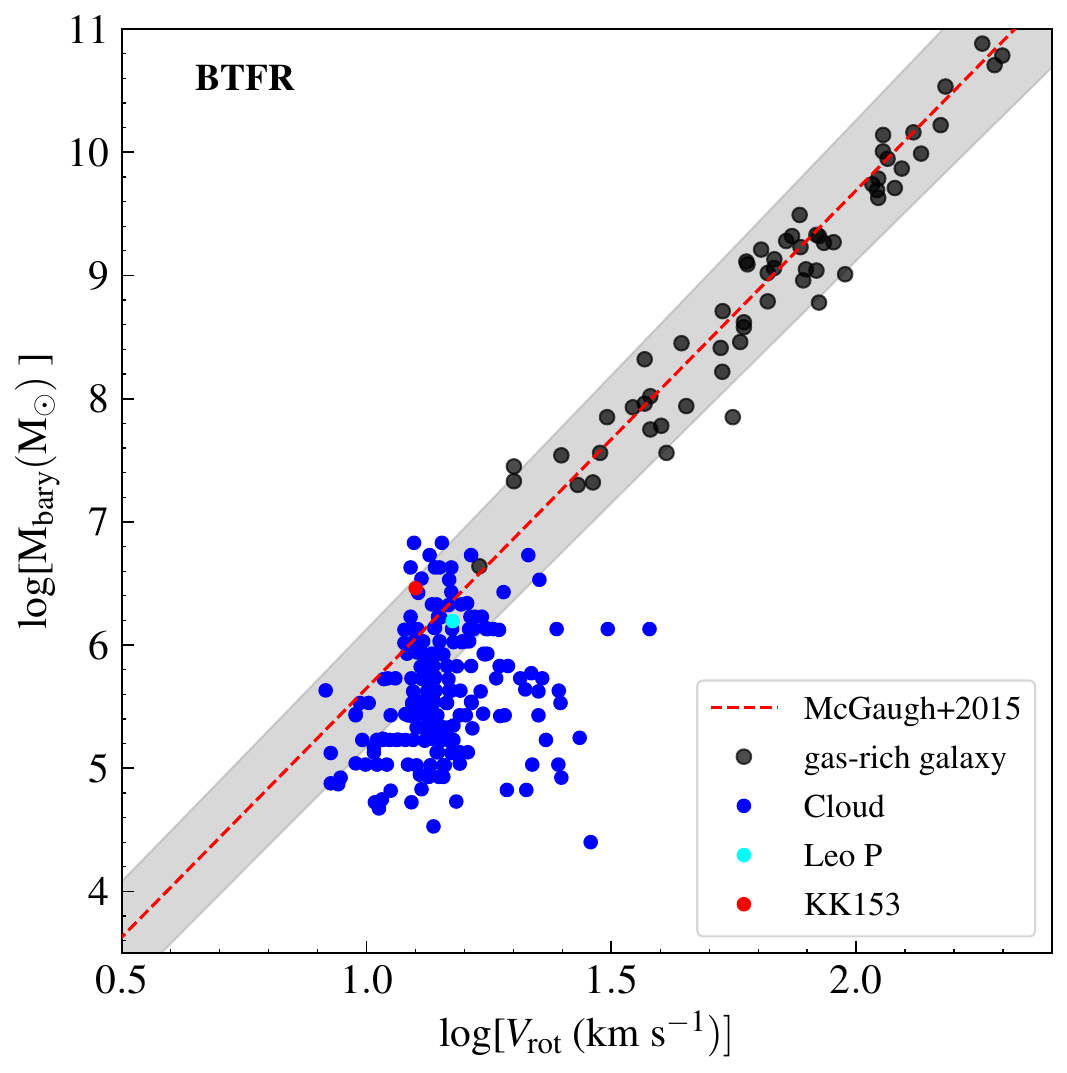}
\vspace{-3mm}
\caption{Left panel: \HI Tully-Fisher relation (HTFR). The black points represent the dwarf galaxies in Local volume \citep{2017AJ....153....6K}. The red dashed line represents a best-fit relation $\rm log(\it M_{\rm HI})=\rm a+b\times log(\it V_{\rm rot})$, where a=2.08 and b =4.24. Right panel: baryonic Tully-Fisher relation (BTFR). The red dashed line represents a best-fit relation $\rm log(\it M_{\rm bar})=\rm a+b\times log(\it V_{\rm rot})$, where a=1.61$\pm$0.18 and b =4.04$\pm$0.09. The gas-dominated galaxies are marked by black points \citep{2015ApJ...802...18M}. The blue solid points represent the identified clouds. In addition to having low baryon masses, Leo P and KK153 also have low dynamical masses.}
\label{fig:TF_DV}
\end{figure}

Distances for the selected CHVCs from the FASHI DR2 source catalog are primarily derived from the Cosmicflows-4 (CF4) distance calculator \citep{2020AJ....159...67K,2024NatAs...8.1610V}.  The Local Group is a gravitationally bound system where internal orbital motions dominate. The CF4 distances are unsuitable for the Local Group because its framework assumes galaxies participate in the Hubble flow with small peculiar velocity perturbations. Here adopting the distance of M31 (0.8 Mpc) for all 185 CHVCs, we derived effective radii ($R_{\rm eff}$), \HI masses ($M_{\rm HI}$), and dynamical masses ($M_{\rm dyn}$). The $R_{\rm eff}$ values range from 0.2 to 3.3 kpc, with a median of 1.0 kpc.  Figure \ref{fig:mass} presents the distributions of $M_{\rm HI}$ and $M_{\rm dyn}$. Both quantities follow a normal distribution rather than a diffuse one, consistent with the interpretation that these CHVCs constitute a coherent population clustered around M31. The $M_{\rm HI}$ values of these CHVCs span 1.9$\times$10$^{5}~\Msun$ to  4.0$\times$10$^{6}~\Msun$,  with a median value of 2.6$\times$10$^{5}~\Msun$. From their mass perspective, these clouds can serve as candidates for gas-rich dwarf galaxies \citep{2008MNRAS.384..535R}, which is associated with the conclusion from the distribution of $D_{\rm maj}$, $D_{\rm maj}$/$D_{\rm min}$, and $S_{\rm sum}$. 

Moreover, the $M_{\rm dyn}$ of these CHVCs range from 1.0$\times$10$^{7}~\Msun$ to  6.3$\times$10$^{8}~\Msun$  with a median value of 4.9$\times$10$^{7}~\Msun$, placing them firmly within the minihalo regime \citep{2010ApJ...708L..22G}. To estimate the baryonic masses ($M_{\rm bar}$), we account for the contribution of helium by scaling the \HI mass by a factor of 1.33, consistent with the primordial helium abundance from Big Bang nucleosynthesis \citep{2020A&A...641A...6P}. In the absence of detected optical counterparts, the stellar mass contribution is negligible, yielding $M_{\rm gas} = 1.33\times M_{\rm HI}$. Figure \ref{fig:dyn_DV} shows $M_{\rm dyn}$ versus $M_{\rm bar}$ for the sample. The $M_{\rm dyn}$/$M_{\rm bar}$ ratios span 17 to 1705, with a median of 130, demonstrating that these CHVCs appear to be strongly dark-matter-dominated. Taken together, their masses, kinematics, and clustering properties identify them as a population of gas-rich, starless minihalos associated with M31.

\begin{figure*}
\centering
\includegraphics[width=0.99\textwidth]{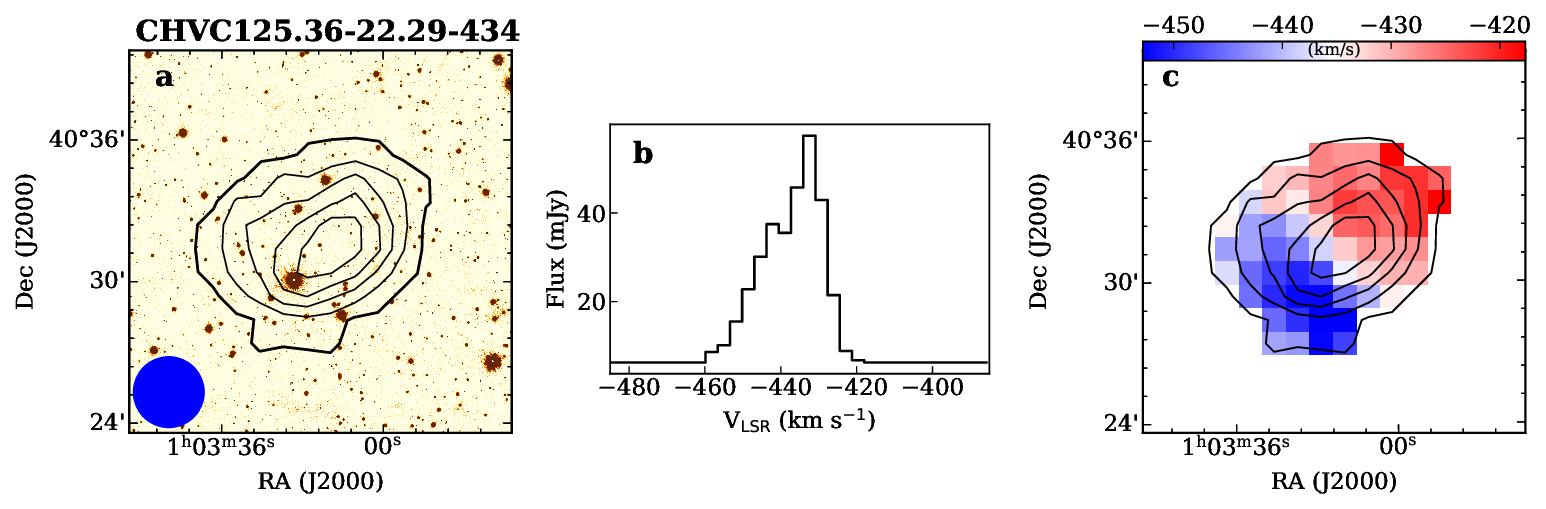}
\vspace{-3mm}
\caption{CHVC125.36-22.29-434. (a), \HI column-density maps from the FAST observation shown in black contours overlaid on the Pan-STARRS1 images in color scale. The black contours begin at 4.2$\times$10$^{17}$ cm$^{-2}$ (3$\sigma$) in steps of 2.0$\times$10$^{18}$ cm$^{-2}$. (b), global \HI profiles shown in a black line.  (c), velocity-field maps in color scale overlaid with its \HI column-density maps in black contours.}
\label{fig:J165_DV}
\end{figure*}

Figure \ref{fig:TF_DV} presents the \HI Tully–Fisher relation (HTFR) and the baryonic Tully–Fisher relation (BTFR) for our sample. \cite{2017AJ....153....6K} established that Local Volume (LV) galaxies with log($M_{\rm HI}$) $>$ 6log$\Msun$ follow a tight linear relation between $M_{\rm HI}$ and $W_{50}$, with the fitting result being largely insensitive to inclination corrections. Although most of our selected CHVCs lie below this mass threshold, their distribution closely traces the extrapolation of the LV galaxy relation.  Adopting the $W_{50}$/2 as the rotational velocity ($V_{\rm rot}$), we find that approximately half of the CHVCs also fall on the BTFR. Moreover, comparing with minihalo models \citep{2013ApJ...768...77A,2002ApJS..143..419S}, which predict median values of \(R_{\mathrm{eff}} \sim 0.7\ \mathrm{kpc}\), \(M_{\mathrm{HI}} \sim 3.0\times10^{5}\ M_{\odot}\), \(N_{\mathrm{HI}} \sim 4.0\times10^{19}\ \mathrm{cm}^{-2}\), and \(M_{\mathrm{dyn}} \sim 3.0\times10^{8}\ M_{\odot}\), we find good overall agreement with our observed median values.
This lends strong support to the interpretation that the CHVCs in our sample are gas-rich, dark-matter-dominated minihalos associated with M31. 

We note that a definitive distance determination for individual CHVCs remains a major observational challenge. In this study, we assume a distance of \(0.8\ \mathrm{Mpc}\) (M31 distance) for all 185 CHVCs projected near M31.
To assess the impact of distance uncertainties on our derived physical parameters, we explore a plausible distance range of \(0.5\)–\(1.2\,\mathrm{Mpc}\), covering the full radial extent of the M31 halo. Across this entire range, the median \HI masses stay within \((1.0\)–\(5.8)\times10^5\,M_\odot\), effective radii within \(0.6\)–\(1.5\,\mathrm{kpc}\), and dynamical masses within \((2.0\)–\(11.8)\times10^7\,M_\odot\). While the absolute numbers vary by a factor of a few, and  are associated with the minihalo regime, which is predicted by \citep{2013ApJ...768...77A,2002ApJS..143..419S}. For the HTFR, the $M_{\rm HI}$ values of these CHVCs span 0.8$\times$10$^{5}~\Msun$ to 8.3$\times$10$^{6}~\Msun$ using the explored distances. This will only slightly shift their position on HTFR. For the BTFR,  a distance variation therefore shifts the clouds vertically on the BTFR by \(\Delta\log(M_{\mathrm{bary}}) \approx 2\Delta\log(D_{\rm dis})\), up to \(\pm0.4\,\mathrm{dex}\). This is comparable to or slightly larger than the intrinsic scatter of the BTFR (\(\sim0.2\)–\(0.3\,\mathrm{dex}\); \citet{2015ApJ...802...18M}). The finding that approximately half of the CHVCs fall on the BTFR extrapolation is therefore reliable.

All CHVCs in our sample have kinetic temperatures \(>5000\ \mathrm{K}\), placing them firmly in the WNM regime.
The absence of a CNM component suggests that these clouds are unable to cool efficiently enough to form molecular gas and, consequently, stars. This is consistent with their lack of optical counterparts and with theoretical expectations for minihalos exposed to a UV background and/or a hot circumgalactic medium \citep{2003ApJ...587..278W}.
A quantitative comparison between the observed temperature distribution and the predictions of photoionization and heating models would be a natural next step.

In addition, \cite{2013ApJ...768...77A} identified 59 UCHVCs, and suggested that the UCHVCs may be  very low mass galaxies. Here we identified 108 UCHVCs in the 192 CHVCs. However, by comparing with optical image from the Pan-STARRS1 Survey, these CHVCs do not have optical counterparts at the current detection limit and smaller distance. 
This makes CHVC particularly like some dark galaxies. Recently, \cite{2025SciA...11S4057L} identified a CHVC, AC G185.0-11.5, located within the HVC complex AC-I, which is considered as a potential dark galaxy. 
To determine the origin of these UCHVCs, detailed dynamic analysis and deeper optical detection are required. 

In the preliminary dynamic analysis, we found that a UCHVC, CHVC125.36-22.29-434, has a very regular velocity gradient, as shown in Figure \ref{fig:J165_DV}. A regular velocity gradient is considered one of the characteristics of disk galaxies. Due to the absence of tidal tail features detected by this UCHVC, although it is located in the M31 group, it can be ruled out that this feature is due to tidal effects. At the current detection limit of optical sensitivity, we can estimate the maximum stellar mass of CHVC125.36-22.29-434.
The stellar masses ($M_*$) are estimated using the mass-to-light ratio equation  $\log(M_*/L_{\rm g})$=-0.601+1.294($m_{\rm g}^{i}-m_{\rm r}^{i}$), where $L_{\rm g}$ is the g-band luminosity derived from the absolute magnitude \citep{2016AJ....152..177H}. The limiting $m_{\rm g}^{i}$ and $m_{\rm r}^{i}$ are 22.8 and 22.6 magnitude in the g-band and r-band for the Pan-STARRS1 Survey, which are estimated as shown in the observation section. By using the above expressions, we obtained that the limiting $M_*$ is 168~$\Msun$ at distance of 0.8 Mpc. The $M_{\rm HI}$ of CHVC125.36-22.29-434 is 2.0$\times$10$^{5}$ \msol. The $M_{\rm HI}$/$M_*$ value is $>$ 10$^{3}$. Such a low limit stellar mass to a large mass ratio indicates that this UCHVC is almost dark. Hence, these UCHVC provide us with a good sample for selecting dark galaxies. Next, we will systematically analyze and study the kinetics of these UCHVCs.

\section{Conclusions}

We have compiled a catalog of 192 compact high-velocity clouds (CHVCs) from the FASHI DR2 source catalog. Our main findings are as follows.

\begin{enumerate}
\item The vast majority (185 out of 192) of the identified CHVCs are spatially concentrated around M31 and M33. Their LSR velocities are not randomly distributed but strongly concentrated around the systemic velocity of M31, with a median value of -296.1 \kms. This indicates they form a kinematically coherent population associated with the M31 system.

\item The observed median \HI mass (\(2.6\times10^{5}\ M_{\odot}\)), effective radius (1.0 kpc), peak column density (\(1.2\times10^{19}\ \mathrm{cm}^{-2}\)), and dynamical mass (\(4.9\times10^{7}\ M_{\odot}\)) are in good agreement with theoretical predictions for minihalos. All clouds have temperatures \(>5000\ \mathrm{K}\) (warm neutral medium), consistent with being embedded in a hot halo.

\item Half of the CHVCs lie on the extrapolation of the baryonic Tully-Fisher relation defined by local volume galaxies. Their morphological and flux characteristics closely resemble those of known gas-rich dwarf galaxies, with 108 objects meeting the definition of ultra-compact HVCs (UCHVCs), most of which are new discoveries.

\item None of the CHVCs have detectable optical counterparts in Pan-STARRS1 imaging down to a g-band limiting magnitude of 22.8. For one UCHVC exhibiting a regular velocity gradient suggestive of a rotating disk, we place an upper limit on its stellar mass of 168 \msol at 0.8 Mpc. This lack of star formation makes these UCHVCs an ideal sample for identifying and studying candidate dark galaxies.
\end{enumerate}

\normalem
\begin{acknowledgements}
We acknowledge the supports of the National SKA program of China (No. 2025SKA0150101) and the National Key R$\&$D Program of China (Nos. 2025YFE202300 and 2022YFA1602901). This work is also supported by the National Natural Science Foundation of China (Grant Nos. 12373001, 12225303, 12421003), the Chinese Academy of Sciences Project for Young Scientists in Basic Research, grant no. YSBR-063, the Guizhou Provincial Science and Technology Projects (Supported by the Guizhou Provincial Science and Technology Projects (No.QKHFQ[2023]003, No.QKHFQ[2024]001, No.ZDSYS[2023]003, No.QKHJC-ZK[2025]MS015).
\end{acknowledgements}

\bibliographystyle{raa}
\bibliography{sample631}{}

\begin{appendix}

\onecolumn
\section{CHVC catalog}
\label{sec:HIcubes}

\begin{table*}[h!] 
     \scriptsize{
      \caption{Continue: measured and derived properties of CHVCs.}
      \vspace{-8pt}
      \setlength{\tabcolsep}{1.5pt}
      \label{tab:Obs}
      \renewcommand{\arraystretch}{0.7}
      \begin{tabular}{lcccccccccccc}
\noalign{\vspace{0pt}}\hline\hline\noalign{\vspace{0pt}}
Name & FASHI-id &RA+Dec. & $V_{\rm LSR}$ & $D_{\rm maj}*\it D_{\rm min}$ & $W_{50}$  & $S_{\rm sum}$& SNR  & log($N_{\rm HI}$)&  $R_{\rm eff}$ &log$M_{\rm HI}$ & log$M_{\rm dyn}$\\
 CHVC+l+b+$V_{\rm LSR}$&  & [J2000]  & [\kms] &[$\prime\times\prime$]  &  [\kms] &  [mJy$\cdot$ km s$^{-1}$] & & [cm$^{-2}]$ &[kpc] & [$\msol$]&[$\msol$]\\
\noalign{\vspace{0pt}}\hline\noalign{\vspace{0pt}}
CHVC112.34-42.38-236 & 20260002002 & 001815+1948 & -236.1 & 77.2$\times$19.4 & 37.4 & 3.4 & 15.1 & 19.2 & 1.8 & 5.7 & 8.2\\
CHVC109.29-52.94-303 & 20260002012 & 001821+0903 & -303.2 & 21.4$\times$11.2 & 43.6 & 0.5 & 9.3 & 18.5 & 0.7 & 4.9 & 7.9\\
CHVC112.92-43.48-322 & 20260002310 & 002049+1847 & -322.2 & 46.8$\times$26.6 & 32.7 & 3.4 & 19.0 & 19.3 & 1.6 & 5.7 & 8.0\\
CHVC111.23-50.09-262 & 20260002313 & 002050+1204 & -261.7 & 22.0$\times$16.0 & 22.3 & 0.8 & 12.8 & 18.9 & 0.9 & 5.1 & 7.4\\
CHVC113.93-41.89-270 & 20260002608 & 002252+2028 & -270.0 & 67.4$\times$32.6 & 35.1 & 6.7 & 32.5 & 19.6 & 2.2 & 6.0 & 8.2\\
CHVC112.44-50.03-360 & 20260002734 & 002356+1216 & -360.4 & 28.0$\times$22.8 & 27.2 & 1.7 & 14.0 & 19.1 & 1.2 & 5.4 & 7.7\\
CHVC116.36-29.69-390 & 20260002778 & 002414+3250 & -390.4 & 20.0$\times$14.4 & 28.5 & 0.9 & 11.3 & 18.7 & 0.8 & 5.1 & 7.6\\
CHVC113.37-48.25-221 & 20260002891 & 002514+1407 & -221.4 & 20.0$\times$16.6 & 24.0 & 0.8 & 10.5 & 18.8 & 0.8 & 5.1 & 7.5\\
CHVC114.75-42.24-313 & 20260002940 & 002539+2013 & -312.6 & 23.2$\times$14.0 & 31.0 & 0.7 & 10.1 & 18.7 & 0.8 & 5.0 & 7.7\\
CHVC112.39-53.56-332 & 20260003011 & 002611+0847 & -332.0 & 56.8$\times$21.0 & 41.2 & 2.7 & 15.6 & 19.1 & 1.6 & 5.6 & 8.2\\
CHVC102.47-72.01-305 & 20260003020 & 002615-1008 & -305.0 & 76.8$\times$56.4 & 42.8 & 26.7 & 40.2 & 20.2 & 3.1 & 6.6 & 8.5\\
CHVC111.30-59.20-377 & 20260003201 & 002742+0308 & -376.8 & 58.8$\times$28.2 & 28.2 & 5.3 & 19.6 & 19.5 & 1.9 & 5.9 & 7.9\\
CHVC117.03-32.54-376 & 20260003285 & 002826+3004 & -375.9 & 77.6$\times$59.8 & 25.9 & 17.3 & 47.3 & 20.0 & 3.2 & 6.4 & 8.1\\
CHVC105.79-72.31-269 & 20260003531 & 003034-1008 & -268.8 & 74.0$\times$34.8 & 45.1 & 16.9 & 34.3 & 19.9 & 2.4 & 6.4 & 8.4\\
CHVC113.71-57.56-255 & 20260003647 & 003138+0457 & -254.6 & 46.6$\times$32.4 & 25.4 & 4.3 & 19.4 & 19.4 & 1.8 & 5.8 & 7.8\\
CHVC114.15-56.38-287 & 20260003676 & 003155+0609 & -287.4 & 69.8$\times$63.4 & 27.6 & 21.2 & 53.0 & 20.2 & 3.1 & 6.5 & 8.1\\
CHVC117.53-36.19-297 & 20260003679 & 003157+2628 & -296.6 & 25.2$\times$17.4 & 25.5 & 1.5 & 11.5 & 19.0 & 1.0 & 5.3 & 7.6\\
CHVC115.29-52.44-303 & 20260003741 & 003233+1008 & -303.5 & 13.4$\times$11.4 & 30.5 & 0.3 & 7.7 & 18.3 & 0.6 & 4.6 & 7.5\\
CHVC116.89-43.02-278 & 20260003762 & 003241+1937 & -278.1 & 41.6$\times$32.6 & 26.0 & 2.6 & 14.9 & 19.3 & 1.7 & 5.6 & 7.8\\
CHVC116.05-49.89-315 & 20260003821 & 003316+1244 & -315.3 & 64.8$\times$50.0 & 34.4 & 8.5 & 26.6 & 19.7 & 2.6 & 6.1 & 8.3\\
CHVC114.33-59.43-283 & 20260003893 & 003358+0309 & -283.2 & 36.4$\times$20.0 & 32.7 & 1.7 & 13.0 & 19.0 & 1.3 & 5.4 & 7.9\\
CHVC118.14-39.79-322 & 20260004052 & 003526+2255 & -321.9 & 19.6$\times$16.8 & 28.3 & 1.1 & 12.7 & 18.9 & 0.8 & 5.2 & 7.6\\
CHVC116.29-55.04-329 & 20260004118 & 003606+0738 & -329.1 & 21.6$\times$14.6 & 28.5 & 0.8 & 17.5 & 18.7 & 0.8 & 5.1 & 7.6\\
CHVC119.09-36.17-363 & 20260004297 & 003733+2635 & -363.5 & 20.2$\times$15.6 & 23.2 & 0.8 & 10.1 & 18.9 & 0.8 & 5.1 & 7.4\\
CHVC117.80-54.48-323 & 20260004587 & 003924+0816 & -322.8 & 61.0$\times$40.8 & 35.3 & 4.2 & 18.6 & 19.5 & 2.3 & 5.8 & 8.2\\
CHVC117.60-59.02-269 & 20260004705 & 004026+0344 & -269.0 & 28.8$\times$18.6 & 27.6 & 2.7 & 29.7 & 19.2 & 1.1 & 5.6 & 7.7\\
CHVC120.21-33.22-299 & 20260004789 & 004058+2935 & -298.7 & 11.4$\times$8.8 & 21.2 & 0.2 & 8.9 & 18.3 & 0.5 & 4.5 & 7.1\\
CHVC120.74-23.57-506W & 20260004811 & 004103+3915 & -506.2 & 11.4$\times$11.4 & 34.9 & 4.2 & 46.8 & 19.4 & 0.5 & 5.8 & 7.6\\
CHVC118.86-53.73-325S & 20260004910 & 004141+0904 & -324.9 & 84.8$\times$32.4 & 35.4 & 6.7 & 25.6 & 19.6 & 2.4 & 6.0 & 8.3\\
CHVC118.86-53.84-333 & 20260004916 & 004142+0857 & -332.8 & 114.8$\times$29.8 & 29.9 & 6.7 & 30.8 & 19.6 & 2.7 & 6.0 & 8.2\\
CHVC119.31-56.57-223 & 20260005136 & 004325+0614 & -223.1 & 76.4$\times$59.2 & 32.1 & 10.9 & 25.0 & 19.8 & 3.1 & 6.2 & 8.3\\
CHVC119.34-56.62-224 & 20260005144 & 004330+0611 & -223.6 & 78.4$\times$28.8 & 33.1 & 6.7 & 30.2 & 19.7 & 2.2 & 6.0 & 8.1\\
CHVC121.23-28.54-355 & 20260005203 & 004411+3418 & -355.0 & 27.4$\times$27.4 & 23.9 & 6.6 & 65.2 & 19.7 & 1.3 & 6.0 & 7.6\\
CHVC121.14-53.41-319 & 20260005499 & 004707+0926 & -318.6 & 27.0$\times$20.6 & 26.0 & 1.1 & 13.3 & 19.0 & 1.1 & 5.2 & 7.6\\
CHVC121.23-52.03-344 & 20260005511 & 004711+1049 & -343.6 & 20.2$\times$17.0 & 46.5 & 0.8 & 10.0 & 18.6 & 0.9 & 5.1 & 8.0\\
CHVC121.28-72.94-277 & 20260005762 & 004928-1004 & -277.0 & 79.0$\times$39.6 & 29.5 & 16.9 & 35.2 & 20.0 & 2.6 & 6.4 & 8.1\\
CHVC122.49-54.08-345 & 20260005877 & 005024+0847 & -345.1 & 33.8$\times$20.0 & 31.0 & 1.3 & 14.6 & 19.0 & 1.2 & 5.3 & 7.8\\
CHVC123.06-25.03-273 & 20260006039 & 005203+3750 & -272.7 & 12.0$\times$9.0 & 38.7 & 0.3 & 11.0 & 18.2 & 0.5 & 4.7 & 7.6\\
CHVC123.77-24.89-257 & 20260006427 & 005517+3758 & -257.5 & 12.0$\times$8.6 & 50.0 & 0.4 & 11.9 & 18.2 & 0.5 & 4.8 & 7.8\\
CHVC123.97-29.38-350 & 20260006480 & 005545+3328 & -350.0 & 48.0$\times$20.4 & 16.5 & 2.1 & 18.8 & 19.2 & 1.5 & 5.5 & 7.4\\
CHVC125.64-66.02-351 & 20260006484 & 005551-0310 & -351.1 & 34.4$\times$30.2 & 34.7 & 4.2 & 19.9 & 19.4 & 1.5 & 5.8 & 8.0\\
CHVC124.65-38.43-339S & 20260006643 & 005720+2425 & -339.2 & 50.6$\times$42.8 & 35.0 & 6.9 & 38.0 & 19.7 & 2.2 & 6.0 & 8.2\\
CHVC124.45-23.42-258 & 20260006795 & 005839+3925 & -258.0 & 16.6$\times$12.2 & 28.9 & 0.5 & 15.8 & 18.5 & 0.7 & 4.9 & 7.5\\
CHVC124.84-23.25-278 & 20260006991 & 010031+3935 & -278.5 & 15.2$\times$7.4 & 24.7 & 0.3 & 13.5 & 18.4 & 0.5 & 4.6 & 7.2\\
CHVC126.14-45.40-331 & 20260007054 & 010052+1724 & -331.3 & 36.8$\times$23.8 & 26.6 & 2.1 & 18.2 & 19.3 & 1.4 & 5.5 & 7.8\\
CHVC125.08-24.64-268 & 20260007126 & 010123+3811 & -267.6 & 12.8$\times$11.4 & 42.4 & 0.3 & 9.6 & 18.4 & 0.6 & 4.7 & 7.8\\
CHVC125.33-24.13-235 & 20260007260 & 010240+3841 & -235.0 & 14.2$\times$11.0 & 54.5 & 0.9 & 12.7 & 18.5 & 0.6 & 5.1 & 8.0\\
CHVC125.36-22.29-434 & 20260007324 & 010315+4031 & -434.3 & 20.8$\times$16.0 & 25.1 & 1.4 & 16.6 & 19.1 & 0.8 & 5.3 & 7.5\\
CHVC125.41-22.47-229 & 20260007353 & 010326+4020 & -229.0 & 19.2$\times$18.0 & 20.7 & 0.7 & 10.3 & 18.9 & 0.9 & 5.0 & 7.3\\
CHVC126.76-40.90-290 & 20260007401 & 010355+2152 & -289.5 & 16.0$\times$16.0 & 21.7 & 2.6 & 49.1 & 19.4 & 0.7 & 5.6 & 7.3\\
CHVC126.96-40.54-304S & 20260007478 & 010440+2213 & -304.0 & 21.2$\times$15.4 & 19.0 & 1.4 & 28.7 & 19.2 & 0.8 & 5.3 & 7.2\\
CHVC125.71-22.90-316 & 20260007488 & 010447+3953 & -316.2 & 9.8$\times$8.0 & 20.8 & 0.3 & 14.6 & 18.4 & 0.4 & 4.6 & 7.0\\
CHVC126.00-26.41-209 & 20260007528 & 010504+3623 & -208.7 & 13.2$\times$12.4 & 31.0 & 0.5 & 20.6 & 18.6 & 0.6 & 4.9 & 7.5\\
CHVC128.74-53.98-255 & 20260007540 & 010514+0844 & -255.4 & 68.6$\times$39.0 & 27.6 & 6.9 & 28.5 & 19.6 & 2.4 & 6.0 & 8.0\\
CHVC129.65-53.61-258 & 20260007825 & 010733+0902 & -257.9 & 18.0$\times$11.2 & 19.0 & 0.5 & 12.4 & 18.6 & 0.7 & 4.9 & 7.1\\
CHVC129.58-52.59-268 & 20260007858 & 010749+1004 & -267.6 & 22.8$\times$17.2 & 21.6 & 0.9 & 15.8 & 18.9 & 0.9 & 5.1 & 7.4\\
CHVC133.20-65.79-227 & 20260007912 & 010814-0315 & -227.1 & 29.4$\times$23.4 & 42.2 & 2.2 & 10.9 & 19.1 & 1.2 & 5.5 & 8.1\\
CHVC126.69-25.02-277 & 20260007950 & 010839+3743 & -277.0 & 45.0$\times$16.2 & 29.4 & 2.6 & 25.8 & 19.2 & 1.3 & 5.6 & 7.8\\
CHVC126.80-25.17-272 & 20260008006 & 010908+3734 & -271.6 & 18.2$\times$12.4 & 30.9 & 0.7 & 17.5 & 18.7 & 0.7 & 5.0 & 7.6\\
CHVC133.95-65.99-335 & 20260008036 & 010919-0330 & -335.4 & 29.8$\times$23.2 & 31.1 & 2.1 & 12.6 & 19.2 & 1.2 & 5.5 & 7.8\\
\noalign{\vspace{0pt}}\hline\hline\noalign{\vspace{0pt}}
\end{tabular}}
\end{table*}


\begin{table*} 
     \scriptsize{
      \caption{Continue: measured and derived properties of CHVCs.}
      \vspace{-8pt}
      \setlength{\tabcolsep}{1.5pt}
      \label{tab:Obs}
      \renewcommand{\arraystretch}{0.7}
      \begin{tabular}{lcccccccccccc}
\noalign{\vspace{0pt}}\hline\hline\noalign{\vspace{0pt}}
Name & FASHI-id &RA+Dec. & $V_{\rm LSR}$ & $D_{\rm maj}*\it D_{\rm min}$ & $W_{50}$  & $S_{\rm sum}$& SNR  & log($N_{\rm HI}$)&  $R_{\rm eff}$ &log$M_{\rm HI}$ & log$M_{\rm dyn}$\\
 CHVC+l+b+Vsys&  & [J2000]  & [\kms] &[$\prime\times\prime$]  &  [\kms] &  [mJy$\cdot$ km s$^{-1}$] & & [cm$^{-2}]$ &[kpc] & [$\msol$]&[$\msol$]\\
\noalign{\vspace{0pt}}\hline\noalign{\vspace{0pt}}
CHVC129.80-48.20-286S & 20260008162 & 011020+1425 & -286.4 & 90.2$\times$25.6 & 29.4 & 10.5 & 37.4 & 19.9 & 2.2 & 6.2 & 8.1\\
CHVC131.07-48.12-305 & 20260008607 & 011351+1424 & -305.5 & 19.2$\times$13.6 & 29.8 & 0.7 & 10.0 & 18.7 & 0.8 & 5.0 & 7.6\\
CHVC131.23-47.76-309 & 20260008692 & 011428+1444 & -308.5 & 21.4$\times$16.8 & 37.5 & 1.3 & 15.1 & 18.9 & 0.9 & 5.3 & 7.9\\
CHVC128.47-27.65-291 & 20260008841 & 011522+3457 & -291.3 & 17.4$\times$12.6 & 28.7 & 0.5 & 12.7 & 18.5 & 0.7 & 4.9 & 7.5\\
CHVC131.62-45.40-325 & 20260009043 & 011656+1702 & -324.6 & 35.2$\times$27.0 & 27.5 & 2.6 & 20.6 & 19.2 & 1.4 & 5.6 & 7.8\\
CHVC131.90-46.49-280A & 20260009054 & 011704+1556 & -279.7 & 21.6$\times$20.2 & 24.8 & 1.7 & 12.5 & 19.1 & 1.0 & 5.4 & 7.5\\
CHVC132.10-47.42-316S & 20260009055 & 011704+1459 & -316.3 & 30.6$\times$16.4 & 44.9 & 2.1 & 17.2 & 19.0 & 1.0 & 5.5 & 8.1\\
CHVC130.14-34.97-322 & 20260009165 & 011805+2731 & -322.5 & 26.0$\times$24.2 & 25.7 & 1.7 & 23.9 & 19.2 & 1.2 & 5.4 & 7.7\\
CHVC133.07-48.21-319 & 20260009330 & 011913+1406 & -319.2 & 44.2$\times$28.0 & 26.2 & 4.2 & 29.6 & 19.5 & 1.6 & 5.8 & 7.8\\
CHVC129.87-30.37-333 & 20260009413 & 011943+3207 & -333.3 & 22.6$\times$16.2 & 21.0 & 0.8 & 14.3 & 18.8 & 0.9 & 5.1 & 7.4\\
CHVC133.74-48.67-326 & 20260009556 & 012043+1334 & -326.4 & 20.0$\times$9.0 & 32.2 & 0.7 & 9.8 & 18.6 & 0.6 & 5.0 & 7.6\\
CHVC129.40-25.25-240 & 20260009580 & 012050+3715 & -240.0 & 13.8$\times$13.8 & 27.5 & 1.7 & 46.6 & 19.0 & 0.6 & 5.4 & 7.5\\
CHVC129.54-25.08-238 & 20260009730 & 012136+3724 & -238.3 & 11.8$\times$11.8 & 27.7 & 1.1 & 28.0 & 18.8 & 0.5 & 5.2 & 7.4\\
CHVC134.14-47.85-322 & 20260009816 & 012223+1420 & -322.3 & 29.4$\times$13.4 & 32.9 & 1.0 & 15.1 & 18.8 & 0.9 & 5.2 & 7.8\\
CHVC129.37-20.47-196 & 20260010050 & 012356+4159 & -195.6 & 18.8$\times$9.2 & 16.9 & 0.4 & 14.6 & 18.6 & 0.6 & 4.8 & 7.0\\
CHVC130.04-24.01-251 & 20260010145 & 012437+3824 & -250.9 & 20.4$\times$18.4 & 26.8 & 1.7 & 27.3 & 19.1 & 0.9 & 5.4 & 7.6\\
CHVC135.41-48.80-271 & 20260010206 & 012504+1315 & -270.9 & 60.0$\times$28.4 & 27.7 & 4.2 & 23.4 & 19.5 & 1.9 & 5.8 & 7.9\\
CHVC142.66-62.05-313 & 20260010547 & 012751-0034 & -313.1 & 31.6$\times$13.6 & 24.0 & 1.4 & 12.1 & 19.0 & 1.0 & 5.3 & 7.5\\
CHVC133.90-38.43-260 & 20260010652 & 012853+2338 & -259.8 & 15.6$\times$15.6 & 27.1 & 1.4 & 27.6 & 19.1 & 0.7 & 5.3 & 7.5\\
CHVC136.31-47.02-278 & 20260010678 & 012900+1452 & -278.4 & 51.8$\times$32.4 & 31.3 & 5.3 & 22.6 & 19.5 & 1.9 & 5.9 & 8.0\\
CHVC136.21-46.32-254 & 20260010716 & 012920+1534 & -253.7 & 38.4$\times$35.2 & 25.8 & 3.3 & 16.0 & 19.5 & 1.7 & 5.7 & 7.8\\
CHVC143.55-60.24-291 & 20260011014 & 013142+0100 & -290.9 & 36.4$\times$30.4 & 27.5 & 6.9 & 31.7 & 19.6 & 1.5 & 6.0 & 7.8\\
CHVC137.53-46.88-272 & 20260011104 & 013230+1449 & -271.8 & 47.6$\times$22.4 & 38.9 & 3.4 & 19.1 & 19.2 & 1.5 & 5.7 & 8.1\\
CHVC141.17-48.50-245 & 20260011940 & 014030+1237 & -245.3 & 10.8$\times$10.8 & 28.7 & 0.4 & 12.2 & 18.5 & 0.5 & 4.8 & 7.4\\
CHVC137.46-38.11-274 & 20260012008 & 014107+2321 & -274.5 & 18.8$\times$16.4 & 27.2 & 0.9 & 13.0 & 18.9 & 0.8 & 5.1 & 7.5\\
CHVC159.30-68.85-280 & 20260012063 & 014135-0935 & -280.1 & 28.0$\times$19.0 & 26.0 & 2.7 & 13.9 & 19.3 & 1.1 & 5.6 & 7.6\\
CHVC133.92-21.33-216 & 20260012579 & 014521+4023 & -216.5 & 20.0$\times$15.0 & 30.1 & 1.1 & 15.1 & 18.9 & 0.8 & 5.2 & 7.6\\
CHVC140.29-41.67-297 & 20260012690 & 014605+1920 & -297.4 & 26.2$\times$13.8 & 22.4 & 1.3 & 17.5 & 19.0 & 0.9 & 5.3 & 7.4\\
CHVC145.35-44.93-256 & 20260014027 & 015623+1505 & -256.1 & 24.2$\times$14.8 & 28.8 & 1.1 & 12.7 & 18.9 & 0.9 & 5.2 & 7.6\\
CHVC146.11-44.63-264 & 20260014393 & 015855+1510 & -263.5 & 58.6$\times$30.8 & 28.0 & 8.5 & 32.0 & 19.8 & 2.0 & 6.1 & 8.0\\
CHVC148.36-47.69-235 & 20260014561 & 020007+1145 & -234.9 & 21.6$\times$17.8 & 24.8 & 1.7 & 26.7 & 19.2 & 0.9 & 5.4 & 7.5\\
CHVC146.73-43.18-263 & 20260014970 & 020253+1621 & -263.3 & 25.4$\times$20.6 & 27.9 & 1.3 & 16.6 & 19.0 & 1.1 & 5.3 & 7.7\\
CHVC145.04-31.58-240 & 20260017056 & 021618+2737 & -240.2 & 35.2$\times$23.6 & 25.1 & 1.7 & 20.6 & 19.1 & 1.3 & 5.4 & 7.7\\
CHVC149.17-34.95-247 & 20260018233 & 022417+2308 & -247.4 & 46.2$\times$34.6 & 24.6 & 8.5 & 34.4 & 19.8 & 1.9 & 6.1 & 7.8\\
CHVC155.32-43.80-269 & 20260018351 & 022458+1306 & -269.5 & 40.6$\times$17.2 & 27.7 & 2.1 & 22.0 & 19.2 & 1.2 & 5.5 & 7.7\\
CHVC152.13-37.38-224 & 20260019086 & 022849+1954 & -223.6 & 20.8$\times$17.2 & 26.5 & 1.1 & 17.9 & 18.9 & 0.9 & 5.2 & 7.6\\
CHVC145.99-26.58-244 & 20260019104 & 022855+3153 & -243.8 & 26.0$\times$17.2 & 30.0 & 0.7 & 14.7 & 18.7 & 1.0 & 5.0 & 7.7\\
CHVC152.97-37.83-241 & 20260019343 & 023027+1912 & -241.2 & 17.2$\times$12.6 & 21.0 & 0.5 & 11.6 & 18.7 & 0.7 & 4.9 & 7.2\\
CHVC158.70-44.39-267S & 20260019678 & 023218+1121 & -266.8 & 81.0$\times$36.8 & 29.8 & 13.4 & 39.1 & 20.0 & 2.5 & 6.3 & 8.1\\
CHVC153.50-37.34-255 & 20260019826 & 023259+1925 & -255.2 & 89.4$\times$44.8 & 32.7 & 26.7 & 66.9 & 20.1 & 2.9 & 6.6 & 8.3\\
CHVC156.04-39.78-229 & 20260020229 & 023512+1619 & -229.2 & 70.8$\times$47.6 & 36.3 & 6.7 & 12.0 & 19.6 & 2.7 & 6.0 & 8.3\\
CHVC165.61-49.70-212 & 20260020334 & 023547+0419 & -212.5 & 31.0$\times$17.4 & 29.8 & 2.1 & 20.9 & 19.1 & 1.1 & 5.5 & 7.7\\
CHVC153.99-36.66-255 & 20260020337 & 023550+1950 & -254.9 & 45.4$\times$26.0 & 27.2 & 10.6 & 75.3 & 19.8 & 1.6 & 6.2 & 7.8\\
CHVC167.95-51.41-231 & 20260020399 & 023614+0203 & -230.5 & 45.2$\times$21.4 & 25.4 & 6.7 & 34.1 & 19.7 & 1.4 & 6.0 & 7.7\\
CHVC154.74-36.32-265 & 20260020850 & 023847+1949 & -264.5 & 48.4$\times$39.0 & 31.2 & 10.6 & 63.4 & 19.9 & 2.0 & 6.2 & 8.1\\
CHVC156.10-37.99-272 & 20260020922 & 023911+1750 & -272.2 & 90.6$\times$55.6 & 29.8 & 21.2 & 50.6 & 20.2 & 3.3 & 6.5 & 8.2\\
CHVC177.58-41.53-266 & 20260027368 & 032235+0446 & -266.0 & 45.6$\times$38.2 & 48.9 & 6.7 & 26.8 & 19.4 & 1.9 & 6.0 & 8.4\\
CHVC184.17+42.50+193 & 20260074005 & 090820+3835 & 193.5 & 14.8$\times$14.8 & 23.9 & 5.2 & 112.5 & 19.6 & -- & -- & --\\
CHVC257.54+53.87+189 & 20260097916 & 111209-0000 & 189.1 & 17.0$\times$17.0 & 30.6 & 3.3 & 24.3 & 19.4 & -- & -- & --\\
CHVC282.59+69.42-360 & 20260102880 & 122307+0740 & -359.7 & 4.8$\times$4.8 & 57.4 & 0.1 & 7.0 & 17.6 & -- & -- & --\\
CHVC101.99+60.39+183 & 20260114058 & 140104+5349 & 183.2 & 6.4$\times$6.4 & 34.6 & 1.4 & 88.8 & 18.9 & -- & -- & --\\
CHVC101.79+60.28+193 & 20260120945 & 140202+5351 & 192.8 & 10.8$\times$10.8 & 37.3 & 6.6 & 195.9 & 19.6 & -- & -- & --\\
CHVC66.27+42.95-185 & 20260133332 & 163246+4158 & -184.8 & 15.6$\times$14.0 & 19.9 & 0.5 & 12.1 & 18.8 & -- & -- & --\\
CHVC9.19+22.86-191 & 20260133529 & 164550-0848 & -191.5 & 66.4$\times$40.4 & 32.4 & 6.7 & 16.5 & 19.6 & -- & -- & --\\
CHVC44.53-20.86-215 & 20260161112 & 202624+0016 & -215.4 & 27.8$\times$17.8 & 49.4 & 2.1 & 11.9 & 18.9 & 1.0 & 5.5 & 8.2\\
CHVC48.06-21.64-345 & 20260162778 & 203540+0244 & -344.9 & 16.8$\times$16.6 & 30.1 & 0.8 & 12.7 & 18.8 & 0.8 & 5.1 & 7.6\\
CHVC46.43-24.12-259 & 20260167623 & 204112+0011 & -258.8 & 22.4$\times$19.0 & 32.8 & 1.7 & 15.5 & 19.0 & 1.0 & 5.4 & 7.8\\
CHVC48.16-24.27-235 & 20260167689 & 204455+0128 & -235.1 & 20.8$\times$18.0 & 49.8 & 1.7 & 12.9 & 18.9 & 0.9 & 5.4 & 8.1\\
CHVC49.59-28.68-267 & 20260167749 & 210240+0017 & -266.8 & 17.0$\times$10.4 & 19.6 & 0.8 & 12.7 & 18.9 & 0.6 & 5.1 & 7.1\\
CHVC51.96-31.68-193 & 20260167776 & 211711+0025 & -192.7 & 17.0$\times$17.0 & 29.2 & 3.4 & 29.5 & 19.3 & 0.8 & 5.7 & 7.6\\
\noalign{\vspace{0pt}}\hline\hline\noalign{\vspace{0pt}}
\end{tabular}}
\end{table*}

\begin{table*} 
     \scriptsize{
      \caption{Continue: measured and derived properties of CHVCs.}
      \vspace{-8pt}
      \setlength{\tabcolsep}{1.5pt}
      \label{tab:Obs}
      \renewcommand{\arraystretch}{0.7}
      \begin{tabular}{lcccccccccccc}
\noalign{\vspace{0pt}}\hline\hline\noalign{\vspace{0pt}}
Name & FASHI-id &RA+Dec. & $V_{\rm LSR}$ & $D_{\rm maj}*\it D_{\rm min}$ & $W_{50}$  & $S_{\rm sum}$& SNR  & log($N_{\rm HI}$)&  $R_{\rm eff}$ &log$M_{\rm HI}$ & log$M_{\rm dyn}$\\
 CHVC+l+b+Vsys&  & [J2000]  & [\kms] &[$\prime\times\prime$]  &  [\kms] &  [mJy$\cdot$ km s$^{-1}$] & & [cm$^{-2}]$ &[kpc] & [$\msol$]&[$\msol$]\\
\noalign{\vspace{0pt}}\hline\noalign{\vspace{0pt}}
CHVC56.36-32.88-323 & 20260167968 & 212928+0250 & -323.1 & 20.8$\times$17.6 & 25.0 & 1.7 & 29.9 & 19.0 & 0.9 & 5.4 & 7.5\\
CHVC60.73-32.18-341 & 20260168139 & 213559+0613 & -341.4 & 39.8$\times$24.2 & 26.8 & 3.4 & 24.8 & 19.4 & 1.4 & 5.7 & 7.8\\
CHVC58.63-34.99-342 & 20260168351 & 214044+0308 & -342.3 & 45.2$\times$35.4 & 26.1 & 5.3 & 22.5 & 19.6 & 1.9 & 5.9 & 7.9\\
CHVC58.57-35.03-343 & 20260168473 & 214046+0304 & -342.9 & 22.2$\times$22.2 & 24.2 & 4.2 & 28.3 & 19.5 & 1.0 & 5.8 & 7.5\\
CHVC77.30-20.93-345 & 20260168585 & 214209+2443 & -344.9 & 20.6$\times$12.6 & 24.3 & 0.5 & 10.6 & 18.7 & 0.7 & 4.9 & 7.4\\
CHVC63.65-32.63-323 & 20260168589 & 214334+0752 & -323.5 & 25.8$\times$13.4 & 29.3 & 1.1 & 12.0 & 18.9 & 0.9 & 5.2 & 7.6\\
CHVC63.74-33.79-355 & 20260168614 & 214725+0711 & -354.5 & 22.0$\times$9.0 & 28.1 & 0.4 & 11.0 & 18.5 & 0.7 & 4.8 & 7.5\\
CHVC63.38-34.46-322 & 20260168643 & 214842+0632 & -322.0 & 46.4$\times$19.8 & 36.8 & 2.7 & 17.1 & 19.1 & 1.4 & 5.6 & 8.0\\
CHVC65.04-35.06-320 & 20260168746 & 215405+0711 & -320.2 & 19.8$\times$17.2 & 28.0 & 0.7 & 13.3 & 18.7 & 0.9 & 5.0 & 7.6\\
CHVC71.97-30.07-330 & 20260168779 & 215510+1452 & -330.2 & 26.8$\times$13.2 & 21.0 & 0.8 & 10.3 & 18.9 & 0.9 & 5.1 & 7.4\\
CHVC68.35-37.05-283 & 20260168908 & 220718+0752 & -283.4 & 29.4$\times$17.8 & 26.3 & 1.7 & 19.8 & 19.1 & 1.1 & 5.4 & 7.6\\
CHVC67.93-37.40-281 & 20260168943 & 220724+0722 & -281.3 & 17.4$\times$11.4 & 22.0 & 0.5 & 9.4 & 18.7 & 0.7 & 4.9 & 7.3\\
CHVC67.48-37.75-285 & 20260169312 & 220729+0653 & -285.2 & 18.6$\times$12.0 & 49.3 & 0.5 & 8.0 & 18.5 & 0.7 & 4.9 & 8.0\\
CHVC69.15-36.63-305 & 20260169318 & 220749+0837 & -304.7 & 26.4$\times$16.6 & 29.1 & 1.7 & 17.5 & 19.0 & 1.0 & 5.4 & 7.7\\
CHVC82.16-25.13-335 & 20260169322 & 220910+2436 & -334.7 & 15.6$\times$10.0 & 26.8 & 0.4 & 9.4 & 18.5 & 0.6 & 4.8 & 7.4\\
CHVC78.76-28.90-336 & 20260169336 & 220947+1947 & -335.5 & 10.8$\times$10.8 & 27.8 & 0.7 & 21.3 & 18.7 & 0.5 & 5.0 & 7.4\\
CHVC86.20-21.31-277A & 20260169391 & 221125+2955 & -276.5 & 35.2$\times$23.8 & 24.7 & 2.7 & 21.7 & 19.4 & 1.3 & 5.6 & 7.7\\
CHVC80.61-28.87-389 & 20260169410 & 221456+2051 & -389.3 & 49.4$\times$25.8 & 31.9 & 5.3 & 17.4 & 19.6 & 1.7 & 5.9 & 8.0\\
CHVC84.33-25.49-449 & 20260169469 & 221655+2533 & -449.1 & 19.6$\times$16.8 & 21.6 & 0.8 & 11.7 & 18.9 & 0.8 & 5.1 & 7.4\\
CHVC80.26-30.18-320 & 20260169578 & 221732+1939 & -320.2 & 15.8$\times$15.8 & 26.3 & 1.1 & 28.4 & 19.0 & 0.7 & 5.2 & 7.5\\
CHVC80.42-32.16-380 & 20260169651 & 222318+1813 & -379.9 & 55.8$\times$17.0 & 22.1 & 2.7 & 18.6 & 19.4 & 1.4 & 5.6 & 7.6\\
CHVC68.90-43.52-321 & 20260169669 & 222721+0343 & -321.4 & 49.6$\times$41.8 & 33.3 & 8.5 & 27.5 & 19.7 & 2.1 & 6.1 & 8.1\\
CHVC82.33-32.25-198 & 20260169931 & 222851+1909 & -197.8 & 25.8$\times$23.2 & 29.6 & 0.8 & 9.5 & 18.8 & 1.1 & 5.1 & 7.8\\
CHVC82.14-33.10-382S & 20260170172 & 223033+1823 & -381.9 & 19.8$\times$13.0 & 20.2 & 1.7 & 24.6 & 19.2 & 0.7 & 5.4 & 7.2\\
CHVC84.27-31.39-275 & 20260170258 & 223216+2049 & -274.8 & 34.0$\times$24.0 & 29.5 & 2.1 & 15.9 & 19.2 & 1.3 & 5.5 & 7.8\\
CHVC85.04-30.93-423 & 20260170354 & 223322+2135 & -422.8 & 24.6$\times$15.4 & 25.2 & 1.3 & 14.2 & 19.1 & 0.9 & 5.3 & 7.5\\
CHVC75.27-41.31-251 & 20260170438 & 223447+0835 & -250.8 & 26.4$\times$13.6 & 38.3 & 1.3 & 12.5 & 18.9 & 0.9 & 5.3 & 7.9\\
CHVC88.67-27.21-384 & 20260170494 & 223524+2628 & -383.5 & 13.8$\times$13.8 & 25.3 & 1.1 & 21.7 & 18.7 & 0.6 & 5.2 & 7.4\\
CHVC75.24-42.76-269 & 20260170561 & 223837+0728 & -269.3 & 104.6$\times$33.2 & 27.8 & 10.6 & 29.6 & 20.0 & 2.7 & 6.2 & 8.1\\
CHVC75.57-43.03-264 & 20260170587 & 224004+0726 & -264.2 & 15.4$\times$12.6 & 23.0 & 0.8 & 19.8 & 18.9 & 0.6 & 5.1 & 7.3\\
CHVC84.38-34.76-411 & 20260170820 & 224101+1810 & -411.1 & 10.4$\times$10.2 & 25.9 & 0.3 & 8.3 & 18.4 & 0.5 & 4.7 & 7.3\\
CHVC92.27-24.83-395 & 20260170935 & 224154+3013 & -395.2 & 33.6$\times$24.4 & 22.9 & 2.7 & 19.4 & 19.3 & 1.3 & 5.6 & 7.6\\
CHVC91.79-25.96-398 & 20260170997 & 224256+2901 & -398.1 & 70.0$\times$19.2 & 27.4 & 3.4 & 19.8 & 19.4 & 1.7 & 5.7 & 7.9\\
CHVC95.40-21.92-370 & 20260171067 & 224631+3411 & -370.2 & 46.0$\times$21.4 & 27.2 & 4.2 & 30.0 & 19.5 & 1.5 & 5.8 & 7.8\\
CHVC77.56-43.94-237S & 20260171139 & 224653+0741 & -237.0 & 19.6$\times$14.4 & 25.6 & 1.7 & 20.1 & 19.1 & 0.8 & 5.4 & 7.5\\
CHVC88.72-36.42-396 & 20260171333 & 225715+1845 & -396.3 & 33.0$\times$20.2 & 45.7 & 2.7 & 15.6 & 19.1 & 1.2 & 5.6 & 8.2\\
CHVC73.34-51.45-193 & 20260171358 & 225807+0010 & -193.0 & 17.0$\times$15.0 & 24.9 & 0.8 & 12.1 & 18.8 & 0.7 & 5.1 & 7.4\\
CHVC95.32-27.59-387 & 20260172029 & 225909+2911 & -386.5 & 87.6$\times$33.8 & 32.6 & 8.5 & 25.9 & 19.6 & 2.5 & 6.1 & 8.2\\
CHVC76.64-50.12-253 & 20260172078 & 230053+0232 & -253.5 & 19.4$\times$13.2 & 44.9 & 1.3 & 22.7 & 18.8 & 0.7 & 5.3 & 7.9\\
CHVC76.71-51.32-259 & 20260172147 & 230406+0139 & -259.4 & 19.8$\times$13.2 & 24.3 & 1.3 & 17.6 & 19.0 & 0.8 & 5.3 & 7.4\\
CHVC95.10-31.09-374 & 20260172309 & 230546+2600 & -373.6 & 32.2$\times$17.6 & 26.5 & 1.3 & 13.3 & 19.0 & 1.1 & 5.3 & 7.7\\
CHVC77.84-53.23-283 & 20260172547 & 231105+0037 & -282.8 & 41.8$\times$39.8 & 38.1 & 13.4 & 55.1 & 19.8 & 1.9 & 6.3 & 8.2\\
CHVC88.16-49.50-290 & 20260172715 & 232343+0721 & -289.9 & 43.8$\times$20.4 & 62.2 & 6.7 & 34.8 & 19.3 & 1.4 & 6.0 & 8.5\\
CHVC98.74-34.17-397 & 20260173123 & 232353+2433 & -397.0 & 21.8$\times$17.4 & 16.9 & 0.7 & 10.8 & 18.9 & 0.9 & 5.0 & 7.2\\
CHVC97.78-40.85-386 & 20260174313 & 233221+1807 & -386.0 & 14.2$\times$12.4 & 17.7 & 0.4 & 8.5 & 18.6 & 0.6 & 4.8 & 7.1\\
CHVC102.76-32.76-436 & 20260174330 & 233523+2706 & -435.7 & 73.6$\times$45.8 & 25.0 & 5.3 & 15.4 & 19.7 & 2.7 & 5.9 & 8.0\\
CHVC99.32-40.32-422 & 20260175199 & 233602+1904 & -422.1 & 31.8$\times$16.6 & 31.9 & 1.3 & 13.1 & 19.0 & 1.1 & 5.3 & 7.8\\
CHVC99.82-41.73-373 & 20260175495 & 233942+1754 & -373.3 & 11.8$\times$11.8 & 26.3 & 0.8 & 14.9 & 18.7 & 0.5 & 5.1 & 7.3\\
CHVC104.59-32.05-402 & 20260175545 & 234055+2817 & -401.8 & 34.6$\times$21.0 & 24.9 & 2.1 & 12.7 & 19.2 & 1.3 & 5.5 & 7.7\\
CHVC105.63-30.98-401 & 20260175858 & 234314+2935 & -401.0 & 29.6$\times$22.8 & 27.0 & 1.7 & 14.6 & 19.1 & 1.2 & 5.4 & 7.7\\
CHVC104.54-34.42-396 & 20260175976 & 234405+2601 & -396.1 & 40.8$\times$20.8 & 34.2 & 2.1 & 12.3 & 19.1 & 1.4 & 5.5 & 8.0\\
CHVC105.08-34.21-405 & 20260176175 & 234542+2622 & -404.8 & 27.4$\times$15.4 & 24.8 & 1.3 & 13.8 & 19.0 & 1.0 & 5.3 & 7.5\\
CHVC105.86-33.86-430 & 20260176256 & 234760+2653 & -429.9 & 55.0$\times$48.8 & 28.3 & 8.3 & 20.9 & 19.7 & 2.4 & 6.1 & 8.1\\
CHVC105.65-36.28-418 & 20260176412 & 235022+2432 & -417.6 & 31.8$\times$18.4 & 25.5 & 1.7 & 11.5 & 19.1 & 1.1 & 5.4 & 7.6\\
CHVC85.27-66.72-283 & 20260176638 & 235458-0804 & -283.0 & 29.2$\times$17.6 & 32.4 & 5.3 & 33.8 & 19.5 & 1.1 & 5.9 & 7.8\\
CHVC88.42-65.18-298 & 20260176879 & 235605-0605 & -297.7 & 18.2$\times$12.8 & 19.4 & 1.7 & 22.9 & 19.2 & 0.7 & 5.4 & 7.2\\
CHVC107.49-35.93-337 & 20260177340 & 235615+2516 & -336.9 & 52.0$\times$29.8 & 30.1 & 5.3 & 26.2 & 19.5 & 1.8 & 5.9 & 8.0\\
CHVC90.82-63.96-316 & 20260177447 & 235717-0431 & -316.4 & 49.2$\times$40.8 & 24.6 & 21.2 & 73.1 & 20.2 & 2.1 & 6.5 & 7.9\\
CHVC111.35-24.22-263 & 20260177471 & 235807+3726 & -263.3 & 13.8$\times$12.8 & 25.7 & 0.4 & 16.1 & 18.5 & 0.6 & 4.8 & 7.4\\
CHVC109.69-30.66-388 & 20260177564 & 235821+3050 & -387.9 & 50.6$\times$39.6 & 28.7 & 4.2 & 16.9 & 19.4 & 2.1 & 5.8 & 8.0\\
CHVC111.12-25.89-262 & 20260177635 & 235859+3546 & -261.6 & 15.2$\times$11.6 & 21.5 & 0.3 & 8.8 & 18.5 & 0.6 & 4.6 & 7.2\\
CHVC107.96-37.07-386 & 20260177656 & 235909+2416 & -386.3 & 13.4$\times$10.4 & 27.0 & 0.5 & 9.8 & 18.5 & 0.5 & 4.9 & 7.4\\
\noalign{\vspace{0pt}}\hline\hline\noalign{\vspace{0pt}}
\end{tabular}}
\end{table*}

\end{appendix}

\end{document}